\documentclass[openacc]{rstransa}

\usepackage{amsmath}
\usepackage{graphicx}
\usepackage{soul}
\usepackage[binary-units]{siunitx}
\usepackage{mathtools}
\usepackage{pgfplots}
	\pgfrealjobname{RS_paper_ODE}
	\usepgfplotslibrary{fillbetween}
\usepackage{multirow}
\usepackage[square]{natbib}
\setcitestyle{numbers}

\renewcommand\hl[1]{#1}

\jname{rsta}
\Journal{Phil. Trans. R. Soc}

\begin{document}

\title{Stochastic rounding and reduced-precision fixed-point arithmetic for solving neural ordinary differential equations}

\author{Michael~Hopkins$^{*}$, Mantas~Mikaitis$^{*}$, Dave~R.~Lester and Steve~Furber}

\address{APT research group, School of Computer Science, The University of Manchester}

\subject{Computer Arithmetic, Neural Modelling, Ordinary Differential Equations}


\corres{Michael Hopkins \\
\email{michael.hopkins@manchester.ac.uk}\\
Mantas Mikaitis\\
\email{mantas.mikaitis@manchester.ac.uk} \\ 
$^{*}$ - These authors have contributed to the work equally.}

\begin{abstract}
Although double-precision floating-point arithmetic currently dominates high-performance computing, there is increasing interest in smaller and simpler arithmetic types. The main reasons are potential improvements in energy efficiency and memory footprint and bandwidth. However, simply switching to lower-precision types typically results in increased numerical errors. We investigate approaches to improving the accuracy of \hl{reduced-precision} fixed-point arithmetic types, using examples in an important domain for numerical computation in neuroscience: the solution of Ordinary Differential Equations (ODEs). The Izhikevich neuron model is used to demonstrate that rounding has an important role in producing accurate spike timings from explicit ODE solution algorithms. In particular, fixed-point arithmetic with stochastic rounding consistently results in smaller errors compared to single-precision floating-point and fixed-point arithmetic with round-to-nearest across a range of neuron behaviours and ODE solvers. A computationally much cheaper alternative is also investigated, inspired by the concept of \emph{dither} that is a widely understood mechanism for providing resolution below the least significant bit (LSB) in digital signal processing. These results will have implications for the solution of ODEs in other subject areas, and should also be directly relevant to the huge range of practical problems that are represented by Partial Differential Equations (PDEs).
\end{abstract}

%%%%%%%%%% Insert the texts which can accomdate on firstpage in the tag "fmtext" %%%%%

\begin{fmtext}
\end{fmtext}

\maketitle

\section{Introduction and motivation}
\label{sec:intro}
   
64-bit double-precision floating-point numbers are the accepted standard for numerical computing because they largely shield the user from concerns over numerical range, accuracy and precision. However, they come at a significant cost. Not only does the logic required to perform an arithmetic operation such as multiply-accumulate on this type consume an order of magnitude more silicon area and energy than the equivalent operation on a 32-bit integer or fixed-point type, but each number also requires double the memory storage space and double the memory bandwidth (and hence energy) each time it is accessed or stored. Today energy-efficiency has become a prime consideration in all areas of computing, from embedded systems through smart phones and data centres up to the next-generation exascale supercomputers. The pressure to pursue all avenues to improved energy efficiency is intense, causing all assumptions, including those regarding optimum arithmetic representations, to be questioned.

Nowhere is this pressure felt more strongly than in machine learning, where the recent explosion of interest in deep and convolutional nets has led to the development of highly-effective systems that incur vast numbers of computational operations, but where the requirements for high-precision are limited. As a result, reduced precision types, such as fixed-point real types of various word lengths, are becoming increasingly popular in  architectures developed specifically for machine learning.
Where high throughput of arithmetic operations is required, accuracy is sacrificed by reducing the working numerical type from floating- to fixed-point and the word length from 64- or 32-bit to 16- or even 8-bit precision \cite{Jouppi:2017:IPA:3079856.3080246}.
By reducing the word length, precision is compromised to gain the advantage of a smaller memory footprint and smaller hardware components and buses through which the data travels from memory to the arithmetic units.
On the other hand, changing the numerical type from floating- to fixed-point, significantly reduces the range of representable values, increasing the potential for the under/overflow of the data types. In some tasks these range issues can be dealt with safely by analysing the algorithm in various novel or problem-specific ways, whereas in other tasks it can be a very difficult to generalise effectively across all use cases \cite{KabiB2017AnOverflowFree}.
There are other approaches apart from floating- and fixed-point arithmetics that are worth mentioning: \textit{posit arithmetic} \cite{Gustafson:2017:BFP:3148214.3148220} is a completely new format proposed to replace floats and is based on the principles of \textit{interval arithmetic} and \hl{\textit{tapered arithmetic}} \cite{1671767} (dynamically-sized exponent and significand fields which optimize the relative accuracy of the floating-point format in some specific range of real numbers rather than having the same relative accuracy across the whole range); \textit{bfloat16}, with hardware support in recent Intel processors \cite{intel_bfloat16}, is simply a single precision floating-point type with the 16 bottom bits dropped for hardware and memory efficiency; \textit{flexpoint} \cite{koster2017flexpoint}, an efficient combination of fixed- and floating-point also by Intel; and various approaches to transforming floating-point using, for example the \textit{logarithmic number system} in which multiplication becomes addition \cite{johnson2018rethinking}.

A 32-bit integer adder uses $9\times$ less energy \cite{6757323} and $30\times$ less area \cite{hw-machine-learning-reduced-precision-slides} than a single-precision floating-point adder.
As well as reducing the precision and choosing a numerical type with a smaller area and energy footprint, mixed-precision arithmetic \cite{micikevicius-mixed-precision}, stochastic arithmetic \cite{Gupta:2015:DLL:3045118.3045303} and approximate arithmetic \cite{6569370} have also been explored in the machine learning community with the goal of further reducing the energy and memory requirements of accelerators. Mixed-precision and stochastic techniques help to address precision loss when short word length numerical types are chosen for the inputs to and outputs from a hardware accelerator, and we address these in the present work.
Mixed-precision arithmetic maintains intermediate results in formats different from the input and output data, whereas stochastic arithmetic works by using probabilistic rounding to balance out the errors in conversions from longer to shorter numerical types. These approaches can also be combined. Approximate arithmetic (or more generally approximate computing) is a recent trend which applies the philosophy of adding some level of (application tolerant) error at the software level or inside the arithmetic circuits to reduce energy and area.

Interestingly, similar ideas have been explored in other areas, with one notable example being at the birth of the digital audio revolution where the concept of \emph{dither} became an important contribution to understanding the ultimate low-level resolution of digital systems \cite{vanderkooy1987dither,vanderkooy1989digital}. Practical and theoretical results show that resolution is available well below the least significant bit if such a system is conceived and executed appropriately.  Although the application here is different, the ideas are close enough to be worth mentioning and we show results with dither added at the input to an ODE solver.
Similarly, an \textit{approximate dithering adder}, which alternates between directions of error bounds to compensate for individual errors in the accumulation operation, has been reported \cite{6386754}.

Our main experiments are run on the Spiking Neural Network (SNN) simulation architecture \emph{SpiNNaker}.
SpiNNaker is based on a digital 18-core chip designed primarily to simulate sparsely connected large-scale neural networks with weighted connections and spike-based signalling - a building block of large-scale digital neuromorphic systems that emphasizes flexibility, scaling and low energy use \cite{spinnproject, 10.3389/fnins.2018.00816}.
At the heart of SpiNNaker lie standard ARM968 CPUs. The ARM968 was chosen as it provides an energy-efficient processor core, addressing the need to minimize the power requirements of such a large-scale system. As the core supports only integer arithmetic, fixed-point arithmetic is used to handle real numbers unless the significant speed and energy penalties of software floating-point libraries can be tolerated. While many of the neural models that appear in the neuroscience literature can be simulated with adequate accuracy using fixed-point arithmetic and straightforward algorithms, other models pose challenges and require a more careful approach. In this paper we build upon some previously published work \cite{Hopkins2015} where the Izhikevich neuron model \cite{izhikevic2003simple} was solved using fixed-point arithmetic, and show a number of improvements.

The main contributions of this paper are:

\begin{itemize}
    \item \hl{Improved rounding in fixed-point arithmetic:} GCC implements rounding down in fixed-point arithmetic by default. This includes the rounding of constants in decimal to fixed-point conversion, as well as rounding in multiplication. Together these had a negative impact on accuracy in our earlier work \cite{Hopkins2015}. In order to remedy this, we have implemented our own multiplication routines (including mixed-precision) with rounding control and used those instead of the GCC library routines (Section~\ref{sec:implementation}).

    \item The \hl{accuracy} of ODE solvers for the Izhikevich neuron equations \hl{is} assessed using different rounding routines applied to algorithms that use fixed-point types. This builds on earlier work where the accuracy of various fixed-point neural ODE solvers was investigated for this model \cite{Hopkins2015}, but where the role of rounding was not addressed (Section~\ref{sec:ODEsolvers}). 

    \item Fixed-point ODE solvers are shown to be more robust with \textit{stochastic rounding} than with other rounding algorithms, and are shown experimentally to be more accurate than single-precision floating-point ODE solvers (Section~\ref{sec:ODEsolvers}).
   
    \item 6 bits in the residual and random number are shown to be sufficient for a high-performance stochastic rounding algorithm when used in an ODE solver (Section~\ref{sec:ODEsolvers}\ref{subsec:results}(\ref{subsubsec:comparator})).
    
    \item The application of \textit{dither} at the input to the ODE is shown to be a \hl{simpler and cheaper} but, in many cases, effective alternative to full stochastic rounding (Section~\ref{sec:dither}).
\end{itemize}
   
It should be noted that in this work when we talk about \emph{error} we mean \emph{arithmetic error}. This means the error introduced by imperfections in arithmetic calculations relative to an arithmetic reference, not the error introduced at a higher level by an algorithm relative to a better algorithm. We don't claim that a chosen algorithm is perfect, just that it is a realistic use case on which to test the different arithmetics that are of interest. This point is explained in more depth in Section~\ref{sec:ODEsolvers}\ref{subsec:Izhi}(\ref{subsubsec:AlgoAndArithError}).

\section{Background}
\label{sec:background}

\subsection{Arithmetic types used in this study}
\label{subsec:types}

There are 8 arithmetic types under investigation in this study.  Two of the floating-point types are standardised by the IEEE and widely available on almost every platform, and the other six are defined in the relatively recent ISO standard 18037 outlining C extensions for embedded systems \cite{iso18037} and are implemented by the GCC compiler:

\begin{itemize}
	\item IEEE 64-bit double precision floating-point: \emph{double}
	\item IEEE 32-bit single precision floating-point: \emph{float}
	\item ISO 18037 s16.15 fixed-point: \emph{accum}
	\item ISO 18037 u0.32 fixed-point: \emph{unsigned long fract} or \emph{ulf}
	\item ISO 18037 s0.31 fixed-point: \emph{signed long fract} or \emph{lf}
	\item ISO 18037 s8.7 fixed-point: \emph{short accum}
	\item ISO 18037 u0.16 fixed-point: \emph{unsigned fract} or \emph{uf}
	\item ISO 18037 s0.15 fixed-point: \emph{signed fract} or \emph{f}
\end{itemize}
These types all come with the most common arithmetic operations implemented. The combination of these types and operations (rounded in various ways where appropriate) is the space that we explore in this paper.

\subsection{Fixed-point arithmetic}
\label{subsec:fixed-point-arithmetic-overview}

\begin{figure*}[ht!]
  \centering
  \includegraphics[width=5in]{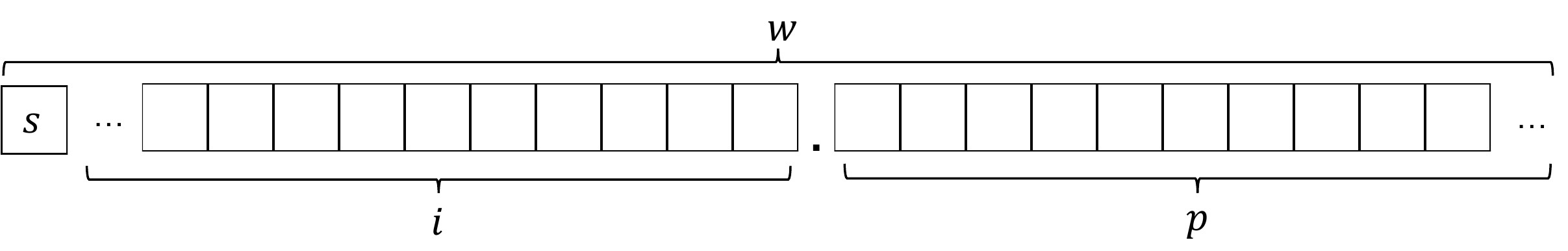}
  \caption{We define a fixed-point number that is made out of three parts: the most significant bit is a sign bit 's', an integer part with $i$ bits and a fractional part with $p$ bits. The \textit{word length} $w=i+p+1$.}
    \label{fig:generalized-fixed-point}
\end{figure*}

A \textbf{generalized fixed-point number} can be represented as shown in Figure~\ref{fig:generalized-fixed-point}.
We define a signed fixed-point number $<s,i,p>$ and an unsigned fixed-point number $<u,i,p>$ of word length $w$ (which usually is $8, 16, 32$ or $64$) where $i$ denotes the number of integer bits, $p$ denotes the number of fractional bits and $s$ is a binary value depending on whether a specific number is signed or unsigned (where $u$ means that there is no sign bit at all and $s$ means there is a sign bit and then we have to use 2's complement in interpreting the numbers).
$p$ also tells us the \textbf{precision} of the representation.
Given a signed fixed-point number $<s,i,p>$, the range of representable values is $[-2^i,2^i-2^{-p}]$.
Whereas, given an unsigned fixed-point number $<u,i,p>$, the range of representable values is $[0,2^i-2^{-p}]$.
To measure the \textbf{accuracy} of a given fixed-point format (or more specifically some function that works in this format and we want to know how well, how \textit{accurately} it performs in a given precision), we define \textit{machine epsilon} $\epsilon=2^{-p}$.
$\epsilon$ gives the smallest positive representable value in a given fixed-point format and therefore represents a \textit{gap} or a \textit{step} between any two neighbouring values in the representable range.
Note that this gap is absolute across the representable range of values and is not relative to the \textit{exponent} as in floating-point or similar representations.
This requires us to consider only absolute error when measuring the accuracy of functions that are implemented in fixed-point representation.
Accuracy is sometimes also measured in terms of LSB - a value represented by the least significant bit in a fixed-point word, which is the same as machine epsilon.
Note that the maximum error of a fixed-point number, when round-to-nearest is used on conversion, is $\frac{\epsilon}{2}$.

Lastly, it is worth noting how to convert a general fixed-point number into a decimal number.
Given a 2's complement fixed-point number of radix 2 (binary vector) $<s,i,p>:sI_{i-1}I_{i-2} \cdot\cdot\cdot I_0.F_{1}F_{2} \cdot\cdot\cdot F_{p}$, if the number is signed the decimal value is given by:

\begin{equation}
value=\sum_{k=0}^{i-1}I_k2^k+\sum_{j=1}^{p}F_j2^{-j}-s2^i.
\end{equation}
Otherwise, if the number is not signed (so bit $s$ becomes integer bit $I_i$) the decimal value is given by:

\begin{equation}
value=\sum_{k=0}^{i}I_k2^k+\sum_{j=1}^{p}F_j2^{-j}.
\end{equation}

\begin{figure}[h!]
  \centering
  \includegraphics[width=3.5in]{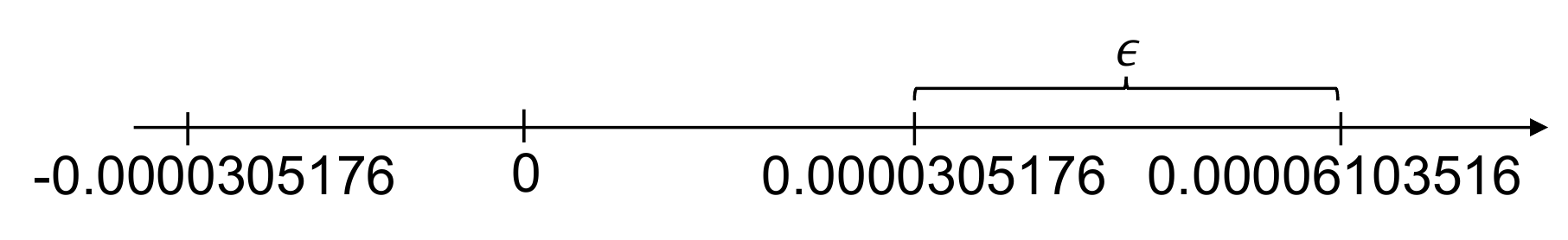}
  \caption{Values close to zero in the \textit{accum} representation.}
  \label{fig:accum-range-example}
\end{figure}

\begin{figure}[h!]
  \centering
  \includegraphics[width=3.5in]{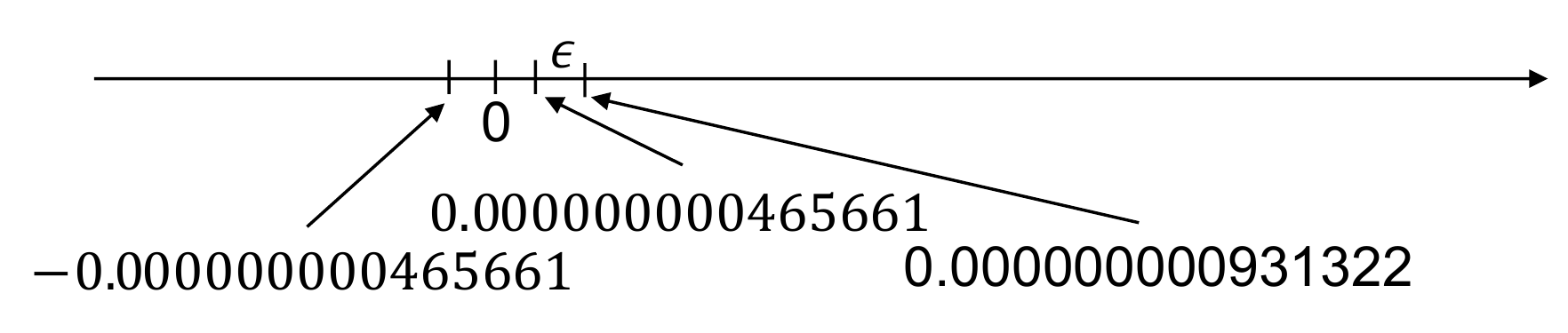}
  \caption{Values close to zero in the \textit{long fract} representation.}
  \label{fig:longfract-range-example}
\end{figure}

SpiNNaker software mostly uses two fixed-point formats available in the GCC implementation: \textit{accum}, which is $<s,16,15>$ and \textit{long fract} which is $<s,0,31>$.
The values close to $0$ for each format are shown in Figures~\ref{fig:accum-range-example} and ~\ref{fig:longfract-range-example}.
The \textit{accum} type has a range of representable values of $[-2^{16}=-65536,2^{16}-2^{-15} = 65535.99996948...]$ with a gap between neighbouring values of $\epsilon=2^{-15} = 0.0000305175...$.
The \textit{long fract} type has a range of $[-1,1-2^{-31}=0.99999999953433...]$ with a gap of $\epsilon=2^{-31}=0.00000000046566...$ between neighbouring values.
As can be seen from the values of machine epsilon, \textit{long fract} is a more precise fixed-point represention.
However, \textit{long fract} has a very small range of representable values compared to \textit{accum}.
Which format should be used depends on the application requirements, such as the required precision and the bounds of all variables.

\begin{figure}[htbp]
  \centering
  \includegraphics[width=3.5in]{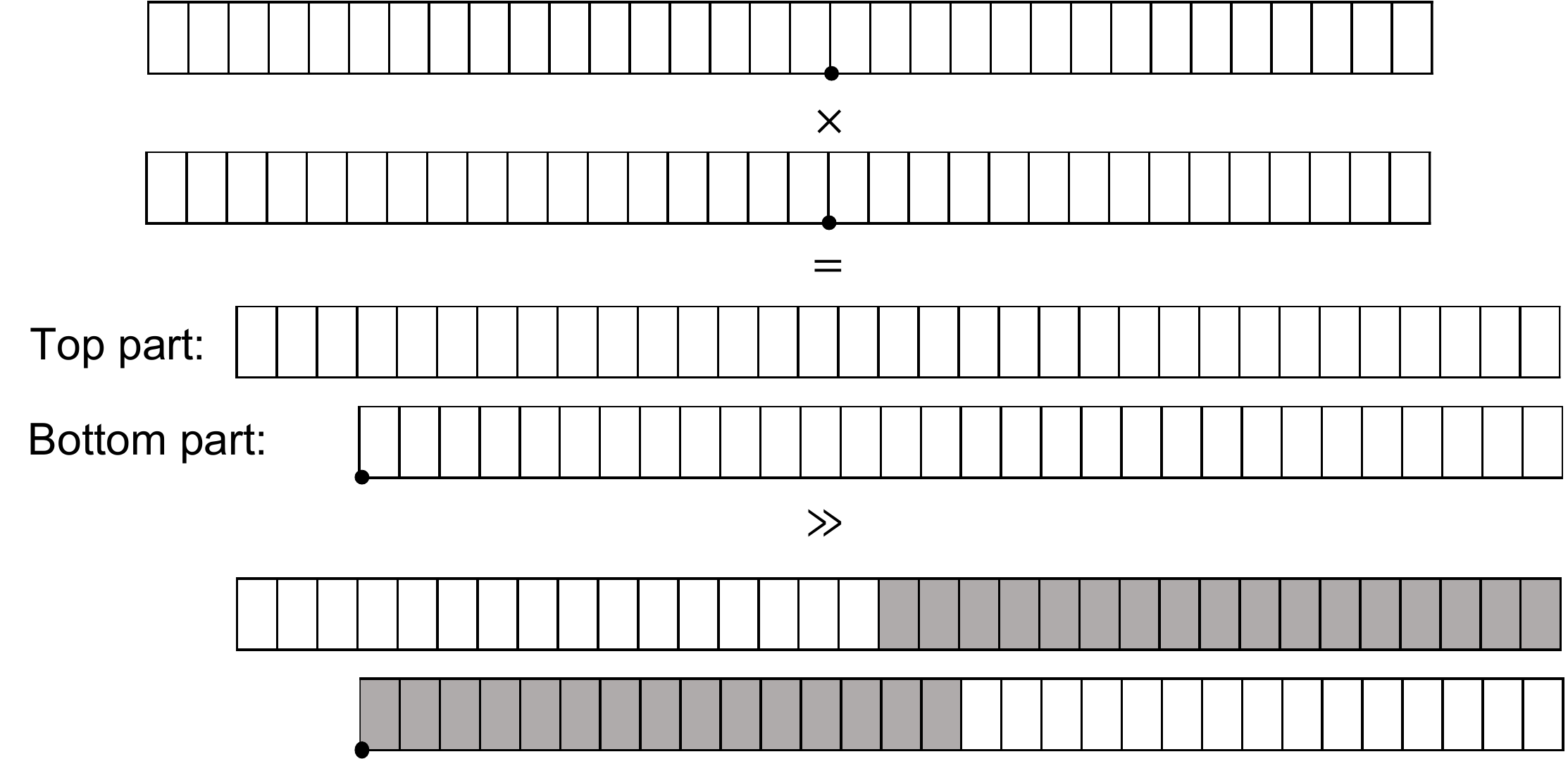}
  \caption{Example showing the multiplication of two \textit{accum} type variables.
    The shaded part is the 32-bit result that has to be extracted, discarding the bottom and top bits.}
  \label{fig:accum-multiplication-example}
\end{figure}

The three main \textbf{arithmetic operations} on fixed-point numbers can be defined as:

\begin{itemize}
    \item Addition: $<s,i,p>+<s,i,p> ~ = ~ <s,i,p>$.
    \item Subtraction: $<s,i,p>-<s,i,p> ~ = ~ <s,i,p>$.
    \item Multiplication: $<s,i_a,p_a>\times<s,i_b,p_b> ~ = ~ <s,i_a+i_b,p_a+p_b>$.
\end{itemize}
Note that $+,-$ and $\times$ denote integer operations available in most of processors' Arithmetic-Logic-Units, including that of the ARM968.
Therefore, for addition and subtraction, no special steps are required (compared to integer operation) when implementing and these operations are exact if there is no underflow or overflow.
However, for multiplication, if the operands $a$ and $b$ have the word lengths $w_a$ and $w_b$, then the result will have the word length of $w_a+w_b-1$.
Therefore after the integer multiplication we have to shift the result right to obtain the result in the same fixed-point representation as the operands (the example in Figure~\ref{fig:accum-multiplication-example} shows this for the \textit{accum} type).
This shifting results in the \textbf{loss of precision} and therefore an appropriate rounding step can be done to minimize the error.

\subsection{Rounding}
\label{subsec:rounding}

Here we will describe two known rounding approaches that we have used to increase the accuracy of the fixed-point  multiplication: \textit{round-to-nearest} (henceforth called \textit{RN}) and \textit{stochastic rounding} (henceforth called \textit{SR}) \cite{Gupta:2015:DLL:3045118.3045303, HOHFELD1992291}; the latter is named stochastic due to its use of random numbers.
Given a real number $x$, and $<s, i, p>$ - an output fixed-point format, \textit{round-to-nearest} is defined as

\begin{align}
\begin{split}
RN(x, <s,i,p>) =
  \left\{
  \begin{matrix*}[l]
    \lfloor x \rfloor, & \text{if} \: \lfloor x \rfloor \leq x < \lfloor x \rfloor + \frac{\epsilon}{2}, \\
    \lfloor x \rfloor + \epsilon, & \text{if} \: \lfloor x \rfloor + \frac{\epsilon}{2} \leq x < \lfloor x \rfloor + \epsilon, \\
  \end{matrix*}
  \right.
\label{equ:rtn}
\end{split}
\end{align}
where $\lfloor x \rfloor$ is the truncation operation (discarding a number of bottom bits and leaving $p$ fractional bits) which returns a number in $<s, i, p>$ format less than or equal to $x$.
Note that we chose to implement \textit{rounding-up} on the tie $x = \lfloor x \rfloor + \frac{\epsilon}{2}$ because this results in a simple rounding routine that requires checking only the \textit{Most Significant Bit (MSB)} of the truncated part to make a decision to round up or down.
Other tie breaking rules such as \textit{round-to-even} can sometimes yield better results, but we prefer the cheaper tie-breaking rule in this work.

Stochastic rounding is similar, but instead of always rounding to the nearest number, the decision about which way to round is non-deterministic and the probability of rounding up is proportional to the residual.
Given all the values as before and, additionally, given a random value $P \in [0, 1)$, drawn from a uniform random number generator, \textit{SR} is defined (\hl{similarly as in} \cite{Gupta:2015:DLL:3045118.3045303}) as

\begin{align}
\begin{split}
SR(x, <s,i,p>) =
 \left\{
  \begin{matrix*}[l]
    \lfloor x \rfloor, & \text{if} \: P \geq \frac{x-\lfloor x \rfloor}{\epsilon}, \\
    \lfloor x \rfloor + \epsilon, & \text{if} \: P < \frac{x-\lfloor x \rfloor}{\epsilon}. \\
  \end{matrix*}
  \right.
  \label{equ:sr}
\end{split}
\end{align}

Lastly, we will denote the cases without rounding as \textit{rounding-down} (henceforth called \textit{RD}) - this is achieved by truncating/shifting(right) a 2's complement binary number.
Note that this results in round-down in 2's complement, while in floating-point arithmetic, which does not use 2's complement, bit truncation results in round towards zero.

\subsection{Floating-point arithmetic}

The IEEE 754-2008 standard radix-2 normalized single-precision floating-point number with a sign bit S, exponent E and an integer significand T, is defined as (\cite[p. 51]{MullerEtAl2018}, slightly modified):

\begin{equation}
	(-1^S) \times 2^{E-127} \times (1+T \cdot 2^{-23}),
\end{equation}
whereas a double-precision number is defined as:

\begin{equation}
	(-1^S) \times 2^{E-1023} \times (1+T \cdot 2^{-52}).
\end{equation}

Table~\ref{table:numerical-types} shows the minimum and maximum values of various floating- and fixed-point numerical types.
It can be seen that fixed-point types have absolute accuracy denoted by the corresponding machine epsilon, which means that any two neighbouring numbers in the representable range have a fixed gap between them.
On the other hand, floating-point types have accuracy relative to the exponent, which means that the gap between any two neighbouring numbers (or a real value of the least significant bit) is relative to the exponent that those numbers have, i.e. the corresponding machine epsilon multiplied by the \num{2} to the power of the biased exponent.
For example, the next number after $0.5 \: (E=126)$ in a single-precision floating-point number is $0.5+2^{-23} \times 2^{-1}$, while the next  number after $1.0 \: (E=127)$ is $1+2^{-23} \times 2^{0}$.
Due to this, the accuracy of floating-point numbers is measured relative to the exponent.
% in ulp which defines the gap between the two floating-point values around the given exact value.
%We refer the readers to \cite{muller:inria-00070503} for a more detailed explanation about ulp(x) for measuring accuracy.

\renewcommand{\arraystretch}{1.15}
\begin{table}[h!]
\centering
	\begin{tabular}{| l || c | c | c | c| c |} 
 	\hline
 	 Property & double precision & single precision & \emph{s16.15} & \emph{u0.32} \\ [0.5ex]
 	\hline\hline
	Accuracy & $2^{-52}$ (Rel.) & $2^{-23}$ (Rel.) & $2^{-15}$ (Abs.) & $2^{-32}$ (Abs.) \\
	Min (Exact) & $2^{-1022}$ & $2^{-126}$ & $2^{-15}$ & $2^{-32}$ \\ 
	Min (Approx.) & $2.225 \times 10^{-308}$ & $1.175 \times 10^{-38}$ & $0.0000305$ & $2.32831 \times 10^{-10}$ \\ 
	Max (Exact) & $(2-2^{-52}) \times 2^{1023}$ & $(2-2^{-23}) \times 2^{127}$ & $2^{16}-2^{-15}$ & $1-2^{-32}$ \\ 
	Max (Approx.) & $1.798 \times 10^{308}$ & $3.403 \times 10^{38}$ & $65535.999969$ & $ 0.999...$ \\ 
 	\hline
	\end{tabular}
	\caption{Some features of floating-point (normalized) and 32-bit fixed-point numerical types.}
	\label{table:numerical-types}
\end{table}

Due to these features of floating-point arithmetic, it depends on the application which types will give more accuracy overall.
For example, if the application works with numbers around \num{1} only, \emph{u0.32} is a more accurate type as it gives smaller steps between adjacent numbers ($2^{-32}$) than does single-precision floating-point ($2^{-23}$).
On the other hand, if the application works with numbers that are as small as $0.001953125 \: (E=117 \text{, which gives a gap between numbers of } 2^{-32})$, single-precision floating-point representation becomes as accurate as \emph{u0.32} and increasingly more accurate as the numbers decrease beyond that point, eventually reaching the accuracy of $2^{-126}$ (normalised) and $2^{-126-23}$ (denormalized) near zero.

Therefore, it is very hard to show analytically what errors a complex algorithm will introduce at various steps of the computation for different numerical types.  More generally, it is important to bear in mind that most of the literature on numerical analysis for computation implicitly assumes constant relative error (as in floating-point) as opposed to constant absolute error (as in fixed-point), and so many results on, for example, convergence and sensitivity of algorithms, need to be considered in this light. In this paper, we address these numerical issues experimentally using a set of numerical types with a chosen set of test cases.
%One approach which may be useful in a limited number of cases such as working with log-likelihoods in the calculation of Bayesian probabilities is to use fixed-point types for the log( $x$ ) of the values needed in the calculations and this then provides constant relative error in the underlying $x$ as well as a huge increase in range (for positive $x$).

\section{Related work}

There is already a body of work analysing reduced precision types and ways to alleviate numerical errors using stochastic rounding, primarily in machine learning research.
In this section we review some of the work that explores fixed-point ODE solvers and stochastic rounding applications in machine learning and neuromorphic hardware.

\subsection{Fixed-point ODE solvers}

The most recent work that explores fixed-point ODE solvers on SpiNNaker \cite{10.3389/fninf.2018.00081}
was published in the middle of our current investigation and exposes some important issues with the default GCC s16.15 fixed-point arithmetic when used with the Izhikevich neuron model.
The authors tested the current sPyNNaker software framework \cite{10.3389/fnins.2018.00816} for simulating the Izhikevich neuron and then demonstrated a method for comparing the network statistics of two forward Euler solvers at small timesteps using a custom fixed-point type of s8.23 for one of the constants. Comparing network behaviour is a valuable development in this area as is their observation about the need for accurate constants in ODE solvers that we learned independently, as described later. However, we do have some comments on their methodology and note a few missing details in their study:

\begin{enumerate}
	
\item Although they mention the value of an absolute reference, they apparently do not use one in their quantitative comparison, and so they compare two algorithms neither of whose accuracy is established.

\item The iterative use of forward Euler solvers with small time steps is far from optimal in terms of the three key measures of accuracy, stability and computational efficiency. Regarding the latter, we estimate that 10x as many operations are required compared to the RK2 solver that they compare against. Given this computational budget, a much more sophisticated solver could be used which would provide higher accuracy and stability than their chosen approach. The use of small time steps also militates against the use of fixed-point types without re-scaling, because of quantisation issues near the bottom of variable ranges that are inevitable with such small time steps.

\item The choice of the s8.23 shifted type to improve the accuracy of the constant 0.04 seems odd when the u0.32 \emph{long fract} type and matching mixed-format arithmetic operations can be used for any constants smaller than 1 and are fully supported by GCC, which could only result in a solver of equal or higher accuracy.

\item Rounding is not explored as a possible improvement in the conversion of constants and arithmetic operations, and neither is the provision of an explicit constant for the nearest value that can be represented in the underlying arithmetic type.

\end{enumerate}
In this study we address all of these issues. 

\subsection{Stochastic rounding}

Randomization of rounding can be traced back to the 1950s \cite{doi:10.1137/1001011}.
Furthermore, a similar idea was also explored in the context of the CESTAC method \cite{sjdc07, alvi08}  where a program was run multiple times, making random perturbations of the last bit of the result of each elementary operation, and then taking a mean to compute the final result (\cite[p. 486]{Higham:2002:ASN} and references therein).

Other studies investigate the effects of probabilistic rounding in backpropagation algorithms \cite{HOHFELD1992291}.
Three different applications are shown with varying degrees of precision in the internal calculations of the backpropagation algorithms.
It is demonstrated that when fewer than 12 bits are used, training of the neural network starts to fail due to weight updates being too small to be captured by limited precision arithmetic, resulting in underflow in most of the updates.
To alleviate this, the authors apply probabilistic rounding in some of the arithmetic operations inside the backpropagation algorithm and show that the neural network can then perform well for word widths as small as 4 bits.
The authors conclude that with probabilistic rounding, the 12-bit version of their system performs as well as a single precision floating-point version.

In a recent paper \cite{Gupta:2015:DLL:3045118.3045303} about SR (and similarly in \cite{Muller2015}), the authors demonstrate how the error resiliency of neural networks used in deep learning can be used to allow reduced precision arithmetic to be employed instead of the standard 32/64-bit types and operations.
The authors describe a simple matrix multiplier array built out of a number of multiply-accumulate units that multiply two 16-bit results and accumulate the results in full-precision (implementing a dot product operation).
Furthermore, the authors show that applying SR when converting the result of the dot product into lower 16-bit precision improves the neural network's training and testing accuracy.
The accuracy with a standard round-to-nearest routine is demonstrated to be very poor while stochastic rounding results are shown to be almost equal to the same neural network that uses single-precision floating-point.
A very important result from this study is that 32-bit fixed-point types can be reduced to 16 bits and maintain neural network performance, provided that SR is applied.
However, the authors did not report the effects of variation in the neural network results due to stochastic rounding - as applications using SR become stochastic themselves it is important to report at least \emph{mean} benchmark results and the \emph{variation} across many trial runs as we have done in this study.

A recent paper from IBM \cite{Wang2018} also explores the use of the 8-bit floating-point type with mixed-precision in various parts of the architecture and stochastic rounding . The authors demonstrate similar performance to the standard 32-bit float type in training neural networks.

In another paper~\cite{10.3389/fnins.2018.00745}, the effects of training recurrent neural networks - used in deep learning applications - on analog Resistive Processing Units (RPUs) were studied.
The authors aim to minimize the analog hardware requirements by looking for a minimum number of bits that the \textit{input arguments} to the analog parts of the circuit can have.
A baseline model with 5-bit input resolution is first demonstrated; it becomes significantly unstable (training error is reported) as network size is increased, compared to a simulation model with single-precision floating point.
The authors then increase the input resolution to \hl{7 bits} and observe a much more regular development of the training error and with lower magnitude at the last training cycle.
Finally, stochastic rounding is applied on these inputs and rounding them to 5 bits again makes the training error almost as stable as the 7-bit version, without the large training error observed in a 5-bit version without stochastic rounding.

In neuromorphic computing SR is used on the recently introduced Intel \textit{Loihi} neuromorphic chip \cite{8259423}.
Here SR is not applied to the ODE of the neuron model as it is in our study, but to the biological learning rule - Spike-Timing-Dependent Plasticity (STDP) - that defines how synapses change with spiking activity.
The implementation of STDP usually requires spike history traces to be maintained for some number of timesteps.
In Loihi, history traces (for each synapse) are held as 7-bit values to minimize memory cost and calculated as $x[t]=\alpha \times x[t-1]+ \beta \times s[t]$, where $\alpha$ is a decay factor which defines how the trace decays back to 0 between spikes and $\beta$ is a value contributed by each spike.
The paper states that these traces are calculated using stochastic rounding.

\section{SR implementation and testing with atomic multiplies}
\label{sec:implementation}

In order to establish the correctness of our rounding mechanisms for fixed-point types, we have carried out a set of tests to assess the distribution of errors from a simple atomic multiply with a wide range of randomly generated operands. There are 5 cases of interest (with 5 equivalent cases for 16-bit types): $\textit{s16.15} \times \textit{s16.15, s16.15} \times \textit{s0.31, s16.15} \times \textit{u0.32, u0.32} \times \textit{u0.32, u0.32} \times \textit{s0.31}$.
The first three of these return a saturated answer in \textit{s16.15} and the last two in \textit{s0.31}, with a sign used because some of the multiplication results in ODE solvers are preceded by a minus to invert the sign.
Each multiplier has an option to do three types of rounding from RD, RN or SR when converting the result to a destination numerical type.
In all cases we take 50,000 random numbers distributed uniformly between a minimum and maximum value. For example, in the cases where both operands are \emph{s16.15} the range is $[-256 , 256]$, or $[-16 , 16]$ for the equivalent 16-bit types, to ensure that the result can be represented. In the other cases the full ranges of the types can safely be used.

The operands are generated in pairs so that the numbers can be represented exactly in the relevant fixed-point type and then copied to the reference type IEEE \emph{double} where the compiler translates the numerical value appropriately. We then multiply them using the default \emph{double} operation and the other multiply operation that under consideration. The fixed-point result is copied back to \emph{double} and the difference between the results is stored and normalised to the LSB of the fixed point result type so that, e.g. if the result is an \emph{s16.15}, the absolute error is multiplied by $2^{15}$. We then calculate summary statistics and plot each set of errors. We will call these \emph{Bit Error Distributions} or \emph{BED}s and they allow the atomic multiplication errors of the different types to be compared directly, as well as to confirm the expected behaviour.

% This routine plots the histograms that are then imported as PDFs above. Turn this on only when you with to replot the diagrams.
% NOTE: The amount of data to process will make pdflatex build quite slow.
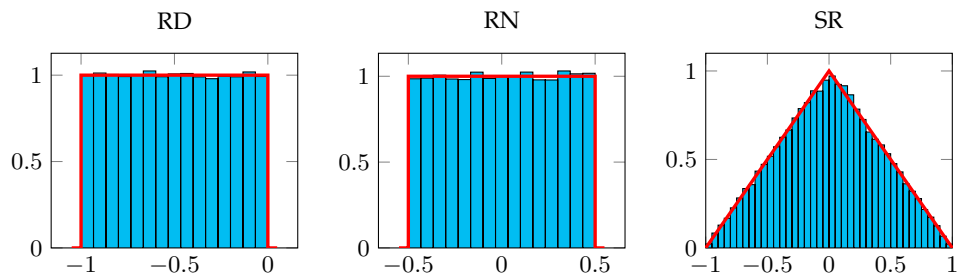
\begin{figure}[h!]
	\begin{center}
    % NOTE: The following avoids the regeneration of the figure and imports the PDF. To regenerate the PDF using the following script
    % run in the terminal 'pdflatex --jobname images/mult-test-histograms RS_paper_ODE'.
	\beginpgfgraphicnamed{mult-test-histograms}
		\begin{tabular}{ccc}
		\begin{tikzpicture}
		  \begin{axis}[width=1.9in, ymin=0,title=\text{RD}]
				\addplot [hist={bins=15, density}, fill=cyan!75, draw=black!50!black]
					table [y index=0] {plotdata/axa-none.txt};
				\addplot+ [line width=1.25pt, red, domain=-1.05:0.05, const plot, mark=none] coordinates {(-1.05,0) (-1,0) (-1, 1) (0, 1) (0, 0) (0.05, 0) };

		  \end{axis}
		\end{tikzpicture}
		&
		\begin{tikzpicture}
		  \begin{axis}[width=1.9in, ymin=0,title=\text{RN}]
				\addplot [hist={bins=15, density}, fill=cyan!75, draw=black!50!black]
					table [y index=0] {plotdata/axa-rtn.txt};
				\addplot+ [line width=1.25pt, red, domain=-1.05:0.05, const plot, mark=none] coordinates {(-0.55,0) (-0.5,0) (-0.5, 1) (0.5, 1) (0.5, 0) (0.55, 0) };
		  \end{axis}
		\end{tikzpicture}
		&
		\begin{tikzpicture}
		  \begin{axis}[xmin=-1.0, xmax=1.0, width=1.9in, ymin=0,title=\text{SR}]
				\addplot [hist={bins=40, density}, fill=cyan!75, draw=black!50!black]
					table [y index=0] {plotdata/axa-sr1.txt};
				\addplot [line width=1.25pt, red, domain=-1.0:1.0, samples=201] {1.0-abs(x)};
		  \end{axis}
		\end{tikzpicture}
		\end{tabular}
	\endpgfgraphicnamed
	\end{center}
\caption{Histograms showing the bit error distribution of \num{50000} random $\emph{s16.15}\times \emph{s16.15}$ operations with different rounding schemes.
RD, RN and SR are round-down, round-to-nearest and stochastic rounding, respectively.}
\label{fig:accum-mult-histograms}
\end{figure}

Figure~\ref{fig:accum-mult-histograms} shows the BEDs of a $\textit{s16.15} \times \textit{s16.15}$ multiplier with the three different rounding modes.
%Table~\ref{table:multiplication-stats} shows statistics for some more multipliers.
These plots and summary statistics calculated on the residuals confirm that in all cases the BEDs are as expected and we can have confidence in their implementation.

\subsection{Pseudo-random number generators}

It is important to note that for \emph{SR} to be competitive in terms of computation time \textit{a very efficient source of high quality pseudo-random numbers is critically important as in many cases it will be drawn on for every multiplication operation}.
We have investigated three sources of uniformly distributed noise in this study, implemented as deterministic pseudo-random number generators (PRNGs) with varying levels of quality and execution speed. 

The reference PRNG is a version of Marsaglia's KISS64 algorithm \cite{marsKISS64}.  This has had several incarnations - we are using David Jones' implementation of what is called KISS99 here \cite{2007TestU01}. It is now the default PRNG on the SpiNNaker system with a long cycle length $\approx2^{123}$ and produces a high-quality random stream of 32-bit integers that satisfy the stringent TESTU01 BigCrush and Dieharder test suites \cite{2007TestU01, dieharder-weblink}. Results have also been checked using faster PRNGs that fail these very challenging tests but which are considered to be of a good basic quality for non-critical applications.  In reducing order of sophistication these are a 33-bit maximum-cycle LFSR algorithm with taps at bits 33 and 20 implemented within the SpiNNaker base runtime library \emph{SARK} \cite{Golomb:1981:SRS:578271} and a very simple Linear Congruential algorithm with a setup recommended by Knuth and Lewis, \emph{ranqd1} (\cite[p. 284]{press1992numerical}). No significant differences in either the mean or the standard deviation (SD) of the results were found in any case tested so far, which is encouraging in terms of showing that SR is insensitive to the choice of PRNG as long as it meets a certain minimum quality standard.

\section{Ordinary Differential Equation solvers}
\label{sec:ODEsolvers}

The solution of Ordinary Differential Equations (ODEs) is an important part of mathematical modelling in computational neuroscience, as well as in a wide variety of other scientific areas such as chemistry (e.g. process design), biology (e.g. drug delivery), physics, astronomy, meteorology, ecology and population modelling.  

In computational neuroscience we are primarily interested in solving ODEs for neuron behaviour though they can also be used for synapses and other more complex elements of the wider connected network such as gap junctions and neurotransmitter concentrations. Many of the neuron models that we are interested in are formally \emph{hybrid systems} in that they exhibit both continuous and discrete behaviour. The continuous behaviour is defined by the time evolution of the internal variables describing the state of the neuron and the discrete behaviour is described by a threshold being crossed, a spike triggered followed by a reset of the internal state and then sometimes a refractory period set in motion whereby the neuron is temporarily unavailable for further state variable evolution.

A number of good references exist regarding the numerical methods required to solve ODEs when analytical solutions are not available, which is usually the case \cite{hall1976modern,lambert1991numerical,dormand1996numerical,butcher2003numerical}.

\subsection{Izhikevich neuron}
\label{subsec:Izhi}

A sophisticated neuron model which provides a wide range of potential neuron behaviours is the one introduced by Eugene Izhikevich \cite{izhikevic2003simple}. In our earlier work \cite{Hopkins2015} we take a detailed look at this neuron model, popular because of its relative simplicity and ability to capture a number of different realistic neuron behaviours. Although we provide a new result in that paper which moves some way towards a closed-form solution, for realistic use cases we still need to solve the ODE using numerical methods. We aimed for a balance between the fidelity of the ODE solution with respect to a reference result and efficiency of computation using fixed-point arithmetic, with our primary aim being real-time operation on a fixed time grid (\SI{1}{\milli\second} or \SI{0.1}{\milli\second}).  The absolute reference used is that of the ODE solver in the Mathematica environment \cite{research2014mathematica} which has many advantages including a wide range of methods that can be chosen adaptively during the solution, a sophisticated threshold-crossing algorithm, the ability to use arbitrary precision arithmetic and meaningful warning messages in the case of issues in the solution that contravene algorithmic assumptions. For the Izhikevich model a well implemented, explicit, variable-timestep solver such as the Cash-Karp variant of the Runge-Kutta 4\textsuperscript{th}/5\textsuperscript{th} order will also provide a good reference as long as it implements an effective threshold crossing algorithm. For other models such as variants of the Hodgkin-Huxley model, it may be that more complex (e.g. implicit) methods will be required to get close to reference results. However, none of these will be efficient enough for real-time operation unless there is a significant reduction in the number of neurons that can be simulated, and so for the SpiNNaker toolchain we chose a variant of the fixed-timestep Runge-Kutta 2\textsuperscript{nd} order solver that was optimised for fixed-point arithmetic to achieve an engineering balance between accuracy and execution speed. 

\subsubsection{\emph{Algorithmic error} and \emph{arithmetic error}}
\label{subsubsec:AlgoAndArithError}

In our earlier work \cite{Hopkins2015} we focused on \emph{algorithmic error} i.e. how closely our chosen algorithm can match output from the reference implementation. This algorithmic error is created by the inability of a less accurate ODE solver method to match the reference result (assuming equivalent arithmetic). As the output in this kind of ODE model is fully described by the time evolution of the state variable(s), it is not easy to generate direct comparisons between these time-referenced vector quantities. There are few single number measures that are feasible to calculate whilst at the same time being useful to the computational neuroscience community. After some experiments we decided on spike lead or lag, i.e. the timing of spikes relative to the reference.  This relies on a relatively simple drift in one direction or the other for spike timings, but in all cases tested so far where the arithmetic error is relatively small it is reliable and easy to understand, measure and communicate.  So we work with the lead or lag of spike times relative to the reference for a set of different inputs (either DC or exponentially falling to approximate a single synapse) and neuron types (three of the most commonly used Izhikevich specifications \cite{izhikevic2003simple}: RS (regular spiking), CH (chattering) and FS (fast spiking)).  We looked at different arithmetics and timesteps and their effect, but these were side issues compared to choosing and implementing an algorithm that balanced execution speed against fidelity to the reference using the fixed-point types defined by the ISO standard \cite{iso18037} at a \SI{1}{\milli\second} timestep. 

As mentioned at the end of the introduction, it should be kept in mind that in this section we are investigating a different matter: that of \emph{arithmetic error}. Now we are in a position where we have chosen an algorithm for the above reasons and our new arithmetic reference is this algorithm calculated in IEEE double precision arithmetic.  So the purpose of this study is different: \emph{\textbf{how closely can other arithmetics come to this arithmetic reference}}?

We provide results below on four different fixed-timestep algorithms in order to demonstrate some generality in the results. In increasing order of complexity and execution time these are two 2\textsuperscript{nd} order and one 3\textsuperscript{rd} order fixed-timestep Runge-Kutta algorithms (see e.g. \cite{butcher2003numerical, dormand1996numerical}) and a variant of the Chan-Tsai algorithm \cite{Chan:2010aa}.  All are implemented using the ESR (explicit solver reduction) approach described in \cite{Hopkins2015} where the combination of ODE definition and solver mechanism are combined and `unrolled' into an equivalent algebraic solution which can then be manipulated algebraically in order to be optimised for speed and accuracy of implementation using the fixed-point types available.  We refer to our earlier work \cite{Hopkins2015} for other possibilities, and it is worth mentioning that we focused primarily there on a \SI{1}{\milli\second} timestep whereas for the current study we are using \SI{0.1}{\milli\second}.  This is for a number of reasons: (1) it is becoming the new informal standard for accurate neural simulation; (2) accuracy is critically dependent upon the timestep chosen and so our solutions will now be significantly closer to the absolute algorithmic reference; and (3) relative differences in arithmetic should therefore be made more apparent.

\subsection{About constants}

The constants that are used in the ODE solvers of the previously explored Izhikevich neuron model \cite{izhikevic2003simple} cannot usually be represented exactly in a digital computer system.
Originally it was left to the compiler to round constants such as \num{0.04} and \num{0.1}.
However, as part of this work we have discovered that the GCC implementation of the fixed-point arithmetic \textit{does not support rounding in decimal to fixed-point conversion} and therefore the specification of constants suffers from truncation error, which can be up to $\epsilon$ for a given target fixed-point type, and for ODE solvers this error will in most cases accumulate.
Additionally we have found that \textit{there is no rounding on most of the common arithmetic operations and conversions involving fixed-point numbers}.
The pragma \textit{FX\_FULL\_PRECISION} defined in the ISO standard \cite{iso18037} is not implemented in the GCC compiler version 6.3.1 that we have used for this work and therefore there is no way to turn on correct rounding.  Experiments on GCC compiler version 7.3.0 confirm that this is still the case.

Our first step was to specify constants in decimal exactly, correctly rounded to the nearest \emph{s16.15} (e.g. explicitly stating $0.040008544921875$ and $0.100006103515625$ as constants instead of \num{0.04} and \num{0.1}) which has reduced the maximum error in the decimal to \emph{s16.15} conversion to $\frac{\epsilon}{2}=\frac{2^{-15}}{2}$.
As a result this has significantly reduced the spike lag previously reported \cite{Hopkins2015}, leading to an understanding that ODE solvers can be extremely sensitive to how the constants are represented.
This was also noted in other work \cite{10.3389/fninf.2018.00081}, and the solution that the authors took there was to add more bits in the fraction of the constants and add some scaling factors to return the final result in \textit{s16.15}.
Following from that, we have taken another step in this direction and have represented all of the constants smaller than \num{1} as \emph{u0.32} instead of \emph{s16.15}, which resulted in a maximum error of $\frac{2^{-32}}{2}$.
The earlier work \cite{10.3389/fninf.2018.00081} used \textit{s8.23} format for these constants, but we think that there is no downside to going all the way to \textit{u0.32} format if the constants are below~\num{1}, and any arithmetic operations involving these constants can output in a different format if more dynamic range and signed values are required.
In order to support this, we have developed libraries for mixed-format multiplication operations, where \emph{s16.15} variables can be multiplied by \emph{u0.32} variables returning an \emph{s16.15} result (as described in detail in Section~\ref{sec:implementation}).
A combination of more accurate representation of the constants, mixed-format multiplication and appropriate rounding has significantly reduced the error in the fixed-point solution of the Izhikevich neuron model from that previously reported \cite{Hopkins2015} even before the \SI{0.1}{\milli\second} timestep is taken into account.
Therefore, we would like to note that all the results below with \textit{round-down} on the multipliers do not reproduce the results from the previous study \cite{Hopkins2015} because these new results are generated with more precise constants, as well as some reordering of arithmetic operations in the originally reported ODE solvers in order to keep the intermediate values in high-precision fixed-point types for as long as possible.

\subsection{Results}
\label{subsec:results}

In this section we present the results from four different ODE solvers with stochastic rounding added on each multiplier at the stage where the multiplier holds the answer in full precision and has to truncate it into the destination format, as described in Section~\ref{sec:background}.
The experiments were run on the SpiNN3 board, which contains 4 SpiNNaker chips with ARM968 cores \cite{spinnproject}.

\subsubsection{Neuron spike timings}
\label{subsubsec:spike-lags}

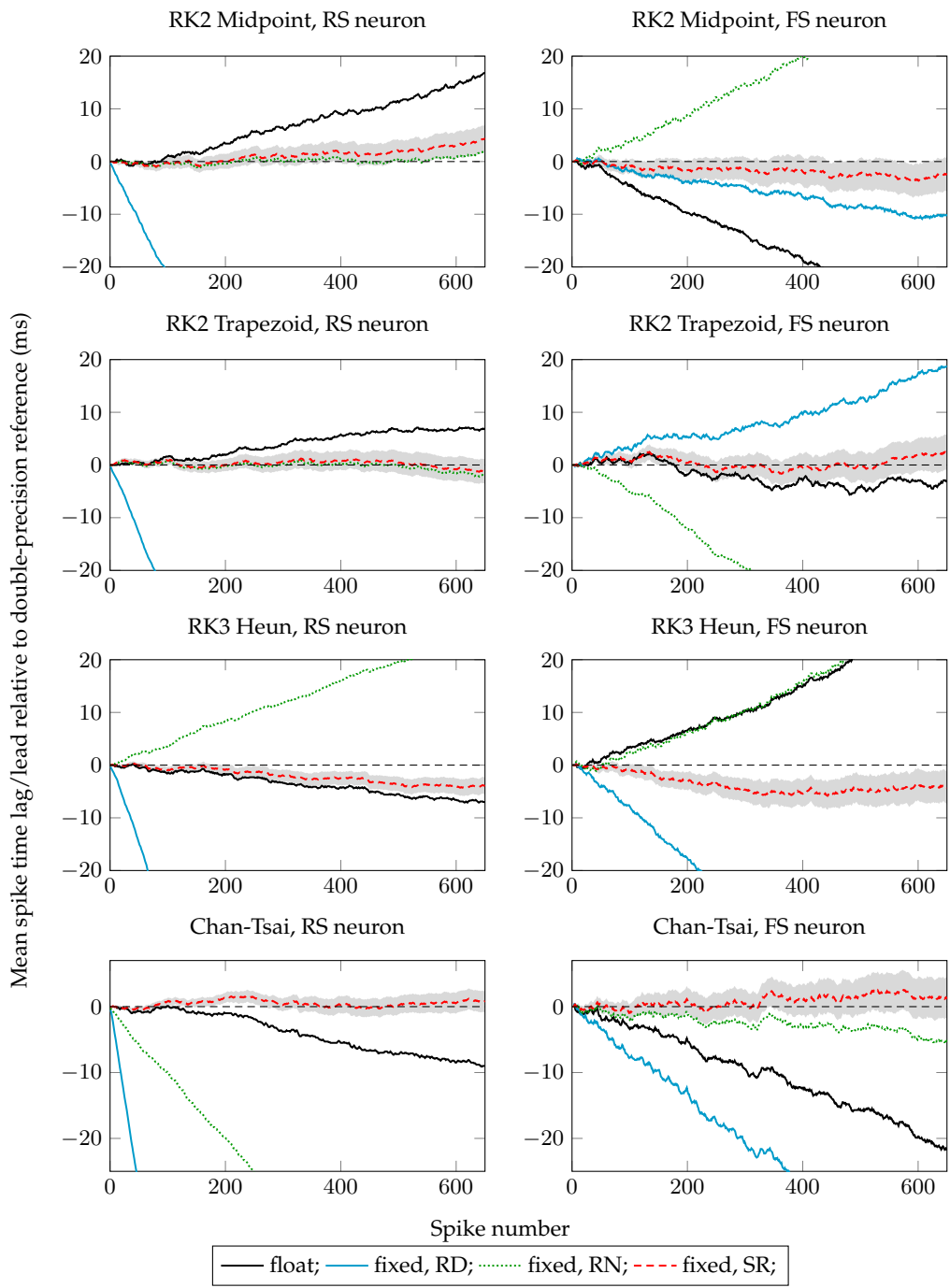
\begin{figure}[ht!]
    \begin{center}
    % NOTE: The following avoids the regeneration of the figure and imports the PDF. To regenerate the PDF using the following script
    % run in the terminal 'pdflatex --jobname images/spike-lag-results RS_paper_ODE'.
	\beginpgfgraphicnamed{spike-lag-results}
    \rotatebox[origin=c]{90}{Mean spike time lag/lead relative to double-precision reference (ms)}
		\begin{tabular}{cc}
        \begin{tikzpicture}
        \begin{axis}[
            title=\text{RK2 Midpoint, RS neuron},
            xmin=0, xmax=650,
            ymin=-20, ymax=20,
            width=2.7in,
            height=1.8in,
            legend columns=-1,
			legend entries={\text{float;}, \text{fixed, RD;}, \text{fixed, RN;}, \text{fixed, SR;}},
			legend to name=legend-spike-lags
        ]
        \addplot+ [line width=0.75pt, mark=none, solid, black] table [x=spike,y=float] {plotdata/lag-RK2-Mid-RS.txt};
        \addplot+ [line width=0.75pt, mark=none, solid, cyan!80!black] table [x=spike,y=accum-nr] {plotdata/lag-RK2-Mid-RS.txt};
        \addplot+ [line width=0.75pt, mark=none, densely dotted, green!60!black] table [x=spike,y=accum-rtn] {plotdata/lag-RK2-Mid-RS.txt};
        \addplot+ [line width=0.75pt, mark=none, densely dashed, red] table [x=spike,y=accum-sr100] {plotdata/lag-RK2-Mid-RS.txt};     
        
        % Draw error bars as a shaded area
        \addplot [name path=upper,draw=none] table[x=spike,y expr=\thisrow{accum-sr100}+\thisrow{sd}] {plotdata/lag-RK2-Mid-RS.txt};
		\addplot [name path=lower,draw=none] table[x=spike,y expr=\thisrow{accum-sr100}-\thisrow{sd}] {plotdata/lag-RK2-Mid-RS.txt};
		\addplot [fill=gray!30] fill between[of=upper and lower];
		
        \addplot+ [color=black, mark=none, densely dashed, domain=0:650]{0}; 
        \end{axis}
        \end{tikzpicture}
        &
        \begin{tikzpicture}
        \begin{axis}[
            title=\text{RK2 Midpoint, FS neuron},
            xmin=0, xmax=650,
            ymin=-20, ymax=20,
            width=2.7in,
            height=1.8in
        ]
        \addplot+ [line width=0.75pt, mark=none, solid, black] table [x=spike,y=float] {plotdata/lag-RK2-Mid-FS.txt};
        \addplot+ [line width=0.75pt, mark=none, solid, cyan!80!black] table [x=spike,y=accum-nr] {plotdata/lag-RK2-Mid-FS.txt};
        \addplot+ [line width=0.75pt, mark=none, densely dotted, green!60!black] table [x=spike,y=accum-rtn] {plotdata/lag-RK2-Mid-FS.txt};
        \addplot+ [line width=0.75pt, mark=none, densely dashed, red] table [x=spike,y=accum-sr100] {plotdata/lag-RK2-Mid-FS.txt};     
        
        % Draw error bars as a shaded area
        \addplot [name path=upper,draw=none] table[x=spike,y expr=\thisrow{accum-sr100}+\thisrow{sd}] {plotdata/lag-RK2-Mid-FS.txt};
		\addplot [name path=lower,draw=none] table[x=spike,y expr=\thisrow{accum-sr100}-\thisrow{sd}] {plotdata/lag-RK2-Mid-FS.txt};
		\addplot [fill=gray!30] fill between[of=upper and lower];
		
        \addplot+ [color=black, mark=none, densely dashed, domain=0:650]{0}; 
        \end{axis}
        \end{tikzpicture}
        \\
        \begin{tikzpicture}
        \begin{axis}[
            title=\text{RK2 Trapezoid, RS neuron},
            xmin=0, xmax=650,
            ymin=-20, ymax=20,
            width=2.7in,
            height=1.8in
        ]
        \addplot+ [line width=0.75pt, mark=none, solid, black] table [x=spike,y=float] {plotdata/lag-RK2-Trap-RS.txt};
        \addplot+ [line width=0.75pt, mark=none, solid, cyan!80!black] table [x=spike,y=accum-nr] {plotdata/lag-RK2-Trap-RS.txt};
        \addplot+ [line width=0.75pt, mark=none, densely dotted, green!60!black] table [x=spike,y=accum-rtn] {plotdata/lag-RK2-Trap-RS.txt};
        \addplot+ [line width=0.75pt, mark=none, densely dashed, red] table [x=spike,y=accum-sr100] {plotdata/lag-RK2-Trap-RS.txt};     
        
        % Draw error bars as a shaded area
        \addplot [name path=upper,draw=none] table[x=spike,y expr=\thisrow{accum-sr100}+\thisrow{sd}] {plotdata/lag-RK2-Trap-RS.txt};
		\addplot [name path=lower,draw=none] table[x=spike,y expr=\thisrow{accum-sr100}-\thisrow{sd}] {plotdata/lag-RK2-Trap-RS.txt};
		\addplot [fill=gray!30] fill between[of=upper and lower];
		
        \addplot+ [color=black, mark=none, densely dashed, domain=0:650]{0}; 
        \end{axis}
        \end{tikzpicture}
        &
    	\begin{tikzpicture}
        \begin{axis}[
            title=\text{RK2 Trapezoid, FS neuron},
            xmin=0, xmax=650,
            ymin=-20, ymax=20,
            width=2.7in,
            height=1.8in
        ]
        \addplot+ [line width=0.75pt, mark=none, solid, black] table [x=spike,y=float] {plotdata/lag-RK2-Trap-FS.txt};
        \addplot+ [line width=0.75pt, mark=none, solid, cyan!80!black] table [x=spike,y=accum-nr] {plotdata/lag-RK2-Trap-FS.txt};
        \addplot+ [line width=0.75pt, mark=none, densely dotted, green!60!black] table [x=spike,y=accum-rtn] {plotdata/lag-RK2-Trap-FS.txt};
        \addplot+ [line width=0.75pt, mark=none, densely dashed, red] table [x=spike,y=accum-sr100] {plotdata/lag-RK2-Trap-FS.txt};     
        
        % Draw error bars as a shaded area
        \addplot [name path=upper,draw=none] table[x=spike,y expr=\thisrow{accum-sr100}+\thisrow{sd}] {plotdata/lag-RK2-Trap-FS.txt};
		\addplot [name path=lower,draw=none] table[x=spike,y expr=\thisrow{accum-sr100}-\thisrow{sd}] {plotdata/lag-RK2-Trap-FS.txt};
		\addplot [fill=gray!30] fill between[of=upper and lower];
		
        \addplot+ [mark=none, densely dashed, domain=0:650]{0}; 
        \end{axis}
        \end{tikzpicture}
        \\
        \begin{tikzpicture}
        \begin{axis}[
            title=\text{RK3 Heun, RS neuron},
            xmin=0, xmax=650,
            ymin=-20, ymax=20,
            width=2.7in,
            height=1.8in
        ]
        \addplot+ [line width=0.75pt, mark=none, solid, black] table [x=spike,y=float] {plotdata/lag-RK3-Heun-RS.txt};
        \addplot+ [line width=0.75pt, mark=none, solid, cyan!80!black] table [x=spike,y=accum-nr] {plotdata/lag-RK3-Heun-RS.txt};
        \addplot+ [line width=0.75pt, mark=none, densely dotted, green!60!black] table [x=spike,y=accum-rtn] {plotdata/lag-RK3-Heun-RS.txt};
        \addplot+ [line width=0.75pt, mark=none, densely dashed, red] table [x=spike,y=accum-sr100] {plotdata/lag-RK3-Heun-RS.txt};     
        
        % Draw error bars as a shaded area
        \addplot [name path=upper,draw=none] table[x=spike,y expr=\thisrow{accum-sr100}+\thisrow{sd}] {plotdata/lag-RK3-Heun-RS.txt};
		\addplot [name path=lower,draw=none] table[x=spike,y expr=\thisrow{accum-sr100}-\thisrow{sd}] {plotdata/lag-RK3-Heun-RS.txt};
		\addplot [fill=gray!30] fill between[of=upper and lower];
		
        \addplot+ [color=black, mark=none, densely dashed, domain=0:650]{0}; 
        \end{axis}
        \end{tikzpicture}
        &
        \begin{tikzpicture}
        \begin{axis}[
            title=\text{RK3 Heun, FS neuron},
            xmin=0, xmax=650,
            ymin=-20, ymax=20,
            width=2.7in,
            height=1.8in
        ]
        \addplot+ [line width=0.75pt, mark=none, solid, black] table [x=spike,y=float] {plotdata/lag-RK3-Heun-FS.txt};
        \addplot+ [line width=0.75pt, mark=none, solid, cyan!80!black] table [x=spike,y=accum-nr] {plotdata/lag-RK3-Heun-FS.txt};
        \addplot+ [line width=0.75pt, mark=none, densely dotted, green!60!black] table [x=spike,y=accum-rtn] {plotdata/lag-RK3-Heun-FS.txt};
        \addplot+ [line width=0.75pt, mark=none, densely dashed, red] table [x=spike,y=accum-sr100] {plotdata/lag-RK3-Heun-FS.txt};     
        
        % Draw error bars as a shaded area
        \addplot [name path=upper,draw=none] table[x=spike,y expr=\thisrow{accum-sr100}+\thisrow{sd}] {plotdata/lag-RK3-Heun-FS.txt};
		\addplot [name path=lower,draw=none] table[x=spike,y expr=\thisrow{accum-sr100}-\thisrow{sd}] {plotdata/lag-RK3-Heun-FS.txt};
		\addplot [fill=gray!30] fill between[of=upper and lower];
		
        \addplot+ [color=black, mark=none, densely dashed, domain=0:650]{0}; 
        \end{axis}
        \end{tikzpicture}
        \\
        \begin{tikzpicture}
        \begin{axis}[
            title=\text{Chan-Tsai, RS neuron},
            xmin=0, xmax=650,
            ymin=-25, ymax=7,
            width=2.7in,
            height=1.8in
        ]
        \addplot+ [line width=0.75pt, mark=none, solid, black] table [x=spike,y=float] {plotdata/lag-Chen-Tsai-RS.txt};
        \addplot+ [line width=0.75pt, mark=none, solid, cyan!80!black] table [x=spike,y=accum-nr] {plotdata/lag-Chen-Tsai-RS.txt};
        \addplot+ [line width=0.75pt, mark=none, densely dotted, green!60!black] table [x=spike,y=accum-rtn] {plotdata/lag-Chen-Tsai-RS.txt};
        \addplot+ [line width=0.75pt, mark=none, densely dashed, red] table [x=spike,y=accum-sr100] {plotdata/lag-Chen-Tsai-RS.txt};     
        
        % Draw error bars as a shaded area
        \addplot [name path=upper,draw=none] table[x=spike,y expr=\thisrow{accum-sr100}+\thisrow{sd}] {plotdata/lag-Chen-Tsai-RS.txt};
		\addplot [name path=lower,draw=none] table[x=spike,y expr=\thisrow{accum-sr100}-\thisrow{sd}] {plotdata/lag-Chen-Tsai-RS.txt};
		\addplot [fill=gray!30] fill between[of=upper and lower];
		
        \addplot+ [color=black, mark=none, densely dashed, domain=0:650]{0}; 
        \end{axis}
        \end{tikzpicture}
        &
        \begin{tikzpicture}
        \begin{axis}[
            title=\text{Chan-Tsai, FS neuron},
            xmin=0, xmax=650,
            ymin=-25, ymax=7,
            width=2.7in,
            height=1.8in
        ]
        \addplot+ [line width=0.75pt, mark=none, solid, black] table [x=spike,y=float] {plotdata/lag-Chen-Tsai-FS.txt};
        \addplot+ [line width=0.75pt, mark=none, solid, cyan!80!black] table [x=spike,y=accum-nr] {plotdata/lag-Chen-Tsai-FS.txt};
        \addplot+ [line width=0.75pt, mark=none, densely dotted, green!60!black] table [x=spike,y=accum-rtn] {plotdata/lag-Chen-Tsai-FS.txt};
        \addplot+ [line width=0.75pt, mark=none, densely dashed, red] table [x=spike,y=accum-sr100] {plotdata/lag-Chen-Tsai-FS.txt};     
        
        % Draw error bars as a shaded area
        \addplot [name path=upper,draw=none] table[x=spike,y expr=\thisrow{accum-sr100}+\thisrow{sd}] {plotdata/lag-Chen-Tsai-FS.txt};
		\addplot [name path=lower,draw=none] table[x=spike,y expr=\thisrow{accum-sr100}-\thisrow{sd}] {plotdata/lag-Chen-Tsai-FS.txt};
		\addplot [fill=gray!30] fill between[of=upper and lower];
		
        \addplot+ [color=black, mark=none, densely dashed, domain=0:650]{0}; 
        \end{axis}
        \end{tikzpicture}
        \\
        \multicolumn{2}{c}{Spike number} \\
    	\multicolumn{2}{c}{\ref*{legend-spike-lags}}
    	\end{tabular}
    \endpgfgraphicnamed
    \end{center}
    \caption{Spike lags of regular and fast spiking Izhikevich neuron models for the DC input test at \SI{0.1}{\milli\second} timestep. Spike lags are computed against a double-precision floating-point reference in each case.
    The SR result is shown as the mean from 100 runs with the shaded area showing the SD.
    A negative value on the Y axis indicates a lead, a positive value indicates a lag.}
    \label{fig:spike-lags0}
\end{figure}

We show results here on the most challenging target problems.  These are a constant DC input of $\approx4.775nA$ for the RS and FS neurons.  The results for exponentially falling input and the CH neuron that were evaluated in our earlier work \cite{Hopkins2015} are essentially perfect in terms of arithmetic error for all the possibilities, therefore we chose not to plot them.
The results for \num{4} different solvers are shown in Figure~\ref{fig:spike-lags0} with the exact spike lags of the 650th spike shown in Table~\ref{table:ode-stats0}.

Because the SR results by definition follow a distribution, we need to show this in the plots.  This has been done by running the ODE solver 100 times with a different random number stream and recording spike times on each run.  We gather a distribution of spike times for each of the first 650 spikes and in Figure~\ref{fig:spike-lags0} show the mean and SD of this distribution. A small number of checks with 1000 repeats show no significant differences in mean or SD from this case.

Clearly the SR results are good in all of these cases compared to the alternatives, and in 7 out of the 8 cases are closest to the arithmetic reference after 650 spikes and look likely to continue this trend. All combinations of four different algorithms with disparate error sensitivities and two different ODE models are shown here, so this bodes well for the robustness of the approach. As a matter of interest, the RK2 Trapezoid algorithm found to produce the most accurate solutions at \SI{1}{\milli\second} without correct rounding of constants or multiplications \cite{Hopkins2015} continues to provide a good performance here in terms of arithmetic error, producing mean spike time errors of only \SI{-1.2}{\milli\second} and \SI{+2.3}{\milli\second} for the RS and FS neuron models after 69 and 165 seconds of simulation time, respectively. Perhaps unsurprisingly, one can see that the SR results look like a random walk.

\begin{table}[h!]
\centering
	\begin{tabular}{| l || c || r | r | r | r |} 
 	\hline
 	Solver & Neuron type & float & fixed, RD & fixed, RN & fixed, SR (Std.Dev.) \\ [0.5ex] 
 	\hline\hline
	\multirow{2}{4em}{RK2 Midpoint} & RS & 16.8 & -131.4 & 1.9 & 4.3 (2.62) \\
														  & FS & -29.7 & -10.0 & 33.7 & -2.3 (3.16) \\
	\hline
	\multirow{2}{4em}{RK2 Trapezoid} & RS & 6.9 & -172.7 & -2.1 & -1.2 (2.30)\\
															& FS & -3.2 & 18.7 & -40.1 & 2.3 (3.33) \\
	\hline
	\multirow{2}{4em}{RK3 Heun} & RS & -7.1 & -206.9 & 26.0 & -4.0 (1.59) \\
													  & FS & 29.4 & -53.6 & 31.4 & -4.4 (3.10) \\
 	\hline
	\multirow{2}{4em}{Chan-Tsai} & RS & -9.0 & -356.3 & -67.9 & 0.8 (1.60) \\
													  & FS & -21.7 & -44.6 & -5.1 & 1.4 (3.10) \\
 	\hline
	\end{tabular}
	\caption{Summary of ODE results: spike lags (ms) of the 650th spike in the DC input test. Positive indicates lag, negative - lead. Fixed, \{RD/RN/SR\} refer to fixed-point arithmetic with round-down, round-to-nearest and stochastic rounding, respectively.}
	\label{table:ode-stats0}
\end{table}

\subsubsection{Evolution of the membrane potential $V$}

\begin{figure}[h!]
    \begin{center}
    % NOTE: The following avoids the regeneration of the figure and imports the PDF. To regenerate the PDF using the following script
    % run in the terminal 'pdflatex --jobname images/membrane-potential-progression RS_paper_ODE'.
	\beginpgfgraphicnamed{membrane-potential-progression}
        \begin{tikzpicture}
        \begin{axis}[
            title=\text{},
            xlabel={Time (ms)},
            ylabel={Membrane potential (mV)},
            xmin=300, xmax=302,
            ymin=-80, ymax=40,
            width=5in,
            height=2in
        ]
        \addplot+ [line width=0.75pt, violet , mark=none] table [x=time,y=mathematica] {plotdata/v-traces-RK2-RS.txt};
        \addplot+ [line width=0.75pt, orange, mark=square] table [x=time,y=double] {plotdata/v-traces-RK2-RS.txt};
        \addplot+ [line width=0.75pt, black, mark=triangle, solid] table [x=time,y=float] {plotdata/v-traces-RK2-RS.txt};
        \addplot+ [line width=0.75pt, cyan!80!black, mark=none] table [x=time,y=accum-nr] {plotdata/v-traces-RK2-RS.txt};
        \addplot+ [line width=0.75pt, green!60!black, densely dotted, mark=x] table [x=time,y=accum-rtn] {plotdata/v-traces-RK2-RS.txt};
        \addplot+ [line width=0.75pt, red, densely dashed, mark=o, solid, error bars/.cd, y dir=both, y explicit] table [x=time,y=accum-sr, y error=sd] {plotdata/v-traces-RK2-RS.txt};
        \addplot [dashed] coordinates {
          (300.5, -80) (300.5, 40)};
        \addplot [dashed] coordinates {
          (301.2, -80) (301.2, 40)};
        \addplot [dashed] coordinates {
          (301.3, -80) (301.3, 40)};
        \legend{\text{Mathematica;}, \text{double;}, \text{float;}, \text{fxp RD;}, \text{fxp RN;}, \text{fxp SR;}}
        \end{axis}
        \end{tikzpicture}
        \endpgfgraphicnamed
    \end{center}
    \caption{The membrane potential of a neuron producing a third spike in the DC current test, using the RK2 Midpoint solver with various arithmetic and rounding formats.
    The temporal difference between the Mathematica and double-precision spike times is \textit{algorithmic error} and the temporal difference between (for example) \textit{fxp RD} and double is \textit{arithmetic error}.}
    \label{fig:membrane-potential-progression}
\end{figure}
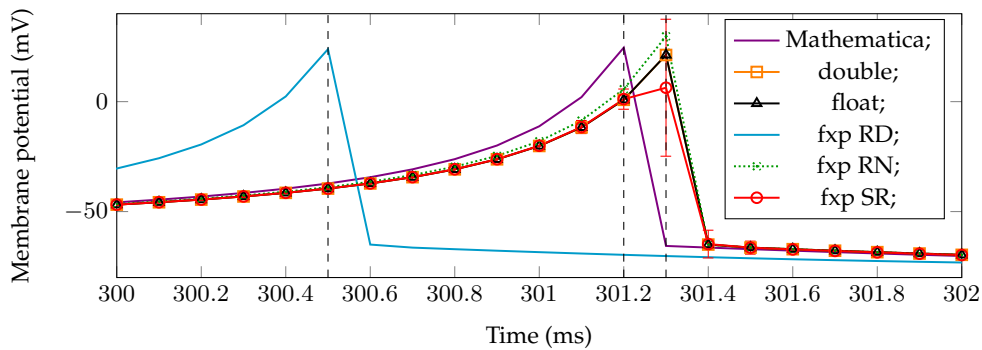

As an illustration of the imperfect evolution of the underlying state variables in the Izhikevich ODE, Figure~\ref{fig:membrane-potential-progression} shows the progression of the $V$ state variable (the membrane potential of the neuron) after \SI{300}{\milli\second} for a variety of solver/arithmetic combinations and with a \SI{0.1}{\milli\second} timestep. One spike and reset event is shown in each case.

The absolute algorithmic reference given by Mathematica is shown in purple. The significant spike time lead given by RD is shown in the light blue line.  The other results show a slight lag relative to the absolute reference, all clustered very close to the arithmetic reference in orange. The SR result shows the mean with SD error bars over 100 random results. The large SD at the spike event, and increased SD near it, are caused by a small number of realisations firing one or two timesteps early or late, inflating the SD sharply as the traces combine the threshold and reset voltages at these time steps. However, the mean behaviour tracks the arithmetic reference very closely.

\subsubsection{The effects of reduced comparator precision in SR}
\label{subsubsec:comparator}

\begin{figure}[h!]
    \begin{center}
    % NOTE: The following avoids the regeneration of the figure and imports the PDF. To regenerate the PDF using the following script
    % run in the terminal 'pdflatex --jobname images/sr-comparator-size RS_paper_ODE'.
	\beginpgfgraphicnamed{sr-comparator-size}
        \begin{tikzpicture}
        \begin{axis}[
            xlabel={Bits in the random number and residual used in stochastic rounding multipliers},
            ylabel={Lag/lead of the 650th spike (ms)},
            xmin=1, xmax=13,
            ymin=-110, ymax=12,
            width=5in,
            height=2.5in,
            legend pos=south east
        ]
        \addplot+ [color=black, mark=x, error bars/.cd, y dir=both, y explicit]
         coordinates {(2, -30.8) +- (2.564310409, 2.564310409)
                             (4, -3.9) +- (1.746511674, 1.746511674)
                             (6, 2.4) +- (2.141287019, 2.141287019)
                             (8, 3.9) +- (2.233492531, 2.233492531)
                             (12, 4.6) +- (2.316957906, 2.316957906)};
        \addplot+ [color=red, mark=square, error bars/.cd, y dir=both, y explicit]
         coordinates {(2, -4.2) +- (3.40227553, 3.40227553)
                             (4, -2.5) +- (3.271463855, 3.271463855)
                             (6, -2.2) +- (3.442036483, 3.442036483)
                             (8, -1.8) +- (3.228655003, 3.228655003)
                             (12, -1.6) +- (3.138648618, 3.138648618)};
        \addplot+ [color=blue, mark=o, error bars/.cd, y dir=both, y explicit]
         coordinates {(2, -46.0) +- (2.952774931, 2.952774931)
                             (4, -9.9) +- (1.404473013, 1.404473013)
                             (6, -3.9) +- (1.792602926, 1.792602926)
                             (8, -1.8) +- (1.964795977, 1.964795977)
                             (12, -1.3) +- (2.095840999, 2.095840999)};
        \addplot+ [color=orange, mark=none, error bars/.cd, y dir=both, y explicit]
         coordinates {(2, 0.4) +- (3.022665055, 3.022665055)
                             (4, 1.5) +- (3.334586885, 3.334586885)
                             (6, 1.4) +- (3.222606229, 3.222606229)
                             (8, 1.8) +- (3.095396386, 3.095396386)
                             (12, 1.9) +- (3.159717894, 3.159717894)};
        \addplot+ [color=cyan!80!black, mark=none, densely dashed, error bars/.cd, y dir=both, y explicit]
         coordinates {(2, -48.5) +- (2.713194658, 2.713194658)
                             (4, -10.6) +- (1.513858872, 1.513858872)
                             (6, -5.8) +- (1.672274203, 1.672274203)
                             (8, -4.3) +- (1.527899132, 1.527899132)
                             (12, -3.8) +- (1.610446954, 1.610446954)};
        \addplot+ [color=violet, mark=*, solid, error bars/.cd, y dir=both, y explicit]
         coordinates {(2, -11.2) +- (3.059424254, 3.059424254)
                             (4, -5.4) +- (3.03445051, 3.03445051)
                             (6, -4.3) +- (3.47792461, 3.47792461)
                             (8, -4.6) +- (3.48852245, 3.48852245)
                             (12, -4.4) +- (3.331650939, 3.331650939)};
        \addplot+ [color=green!60!black, mark=triangle, solid, error bars/.cd, y dir=both, y explicit]
         coordinates {(2, -105.1) +- (2.411793205, 2.411793205)
                             (4, -24.9) +- (2.390676122, 2.390676122)
                             (6, -4.7) +- (1.72347312, 1.72347312)
                             (8, -0.5) +- (1.682425797, 1.682425797)
                             (12, 0.5) +- (1.428977171, 1.428977171)};
        \addplot+ [color=magenta, mark=10-pointed star, solid, error bars/.cd, y dir=both, y explicit]
         coordinates {(2, -15.9) +- (4.107602584, 4.107602584)
                             (4, -2.4) +- (3.34253276, 3.34253276)
                             (6, 1.3) +- (3.541171837, 3.541171837)
                             (8, 1.5) +- (3.352136511, 3.352136511)
                             (12, 1.4) +- (3.2595228, 3.2595228)};
        \legend{\text{RK2 Midpoint, RS}, \text{RK2 Midpoint, FS}, \text{RK2 Trapezoid, RS}, \text{RK2 Trapezoid, FS},
			 						   \text{RK3 Heun, RS}, \text{RK3 Heun, FS}, \text{Chan-Tsai, RS}, \text{Chan-Tsai, FS}}
        \addplot+ [color=black, mark=none, densely dotted, domain=0:13]{0}; 
        \end{axis}
        \end{tikzpicture}
    \endpgfgraphicnamed
    \end{center}
    \caption{Average spike lags of the 650th spike and the standard deviation from 100 runs.
     Four solvers with two different neuron types each are shown for varying numbers of bits in the stochastic rounding comparison step.
     All versions are compared to an equivalent reference solver implemented in double-precision floating-point arithmetic (the dotted straight line at y=0ms).}
    \label{fig:spike-lag1}
\end{figure}
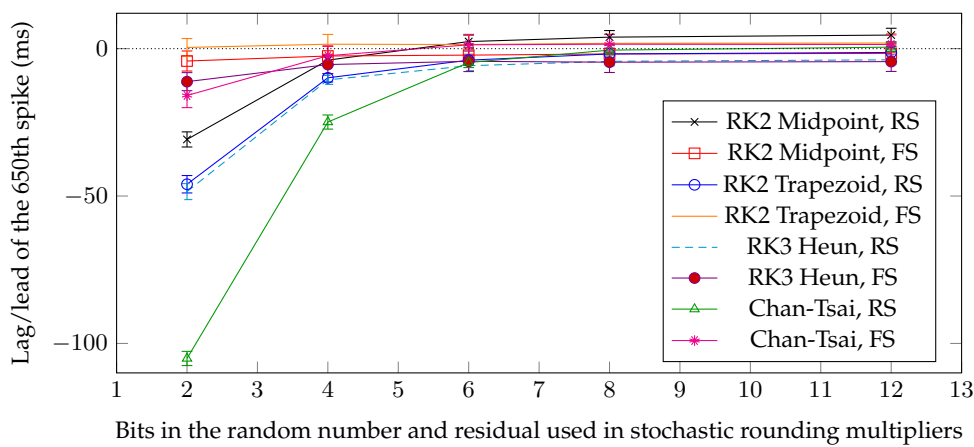

In order to build an efficient SR hardware accelerator it is useful to consider how many bits are required in the rounding calculation.
There are two ways to implement SR: the first is to build a comparator between the residual of the number to be rounded ($\frac{x - \lfloor x \rfloor}{\epsilon}$) and the random number (as in equation~\ref{equ:sr}); the second approach that is strictly equivalent, implemented in an FPGA to optimize the utilization of the Digital Signal Processing (DSP) units \cite{Gupta:2015:DLL:3045118.3045303} is simply to add the random number to the input number and then truncate the fraction.
This reverses equation~\ref{equ:sr} in such a way that the random numbers that are higher than or equal to $1 - (\frac{x-\lfloor x \rfloor}{\epsilon})$ will produce a carry into the result bits (round-up) and the numbers that are lower will not produce a carry (which will result in round-down due to binary truncation).
The first approach has a comparator of the remainder's/random number's width plus an adder after truncation of the remainder, whereas the second approach has only a single adder of the full word width.
Whichever approach is used, the complexity of the adder/comparator can be reduced by defining a minimum number of bits required in the residual $\frac{x-\lfloor x \rfloor}{\epsilon}$ and the random number.

In order to explore this problem in our ODE testbench, we ran a series of ODE solvers on two types of the Izhikevich neuron and measured the effects on the spike lag of different numbers of SR bits.
The results are shown in Figure~\ref{fig:spike-lag1} for the 650th spike.
These results can be compared to the numbers in Table~\ref{table:ode-stats0} where all the available bits were used in SR.
Note that the fixed-point multipliers in the ODE solvers can use 15 or 32 SR bits, depending on the types of the arguments.
It is clear that as we decrease from 12 to 6 bits in the SR, the degradation in quality of the lead or lag relative to the arithmetic reference is negligible.
However, when we decrease the number of bits from 6 to 4 or fewer, the regular spiking neuron starts to perform badly with a very high lag of the 650th spike.
Interestingly, the fast spiking neuron performs quite well even when only 2 SR bits are used.
Given these two tests on the RS and FS neurons, we conclude that 4-bit SR is acceptable, with some degradation in quality of the neuron model, whereas the 6-bit version performs as well as the 12-bit or full remainder length SR version with 15 or 32 bits.

\subsubsection{16-bit arithmetic types}

Using the same constant DC current test applied to a neuron, we evaluate 16-bit numerical types for this problem.  It is a challenging task with such limited precision, but good results would improve memory and energy usage by a substantial further margin.
The results with 16-bit ISO standard fixed-point types in the two second-order ODE solvers are shown in Table~\ref{table:ode-stats-16bit}.
Because most of the variables are now held in \textit{s8.7} numerical type, with 7 bits in the fractional part, even SR performs quite badly.
However, it is clearly better than RD and RN, with RD performing terribly and RN being subject to underflow in one of the variables that causes updates to the main state variable to become 0 and therefore no spikes are produced.
It seems that SR helps the ODE solver to recover from underflow and produce more reasonable answers with a certain amount of lag. 

%It is likely that custom 16-bit data types used within the ESR mechanism for key interim variables (e.g. by scaling a \emph{fract} relative to a known distribution of values that will appear in all possible use cases) would improve this performance further.  For shorter simulations this should be a good tradeoff between performance and accuracy where some spike lag is acceptable and smaller memory footprint and datapath width is a priority.  We aim to investigate this in more detail at a future date and report on the results. It is worth noting that preliminary work using 16-bit types with RN on closed-form solutions for simpler neural models is showing results in some cases identical to \emph{double}.

\begin{table}[h!]
\centering
	\begin{tabular}{| l || c || c | c | c| c |} 
 	\hline
 	Solver & Neuron type & float & fixed, RD & fixed, RN & fixed, SR (Std.Dev.) \\ [0.5ex] 
 	\hline\hline
	\multirow{2}{4em}{RK2 Midpoint} & RS & 16.8 & -21681.4 & - & 889.4 (58.82) \\
														  & FS & -29.7 & -2754.5 & 686.4 & 676.7 (30.67) \\
	\hline
	\multirow{2}{4em}{RK2 Trapezoid} & RS & 6.9 & -22786.2 & - & 363.3 (57.65)\\
															& FS & -4.6 & -2391.2 & 892.8 & 516.7 (27.92) \\
	\hline
	\end{tabular}
	\caption{Summary of the ODE results with 16-bit fixed-point arithmetic compared to a double-precision reference (with the constants representable in 16 bits): spike lags (ms) of the 650th spike in the DC current test. Positive means lag, negative - lead.
	The test cases marked with a dash did not produce any spikes due to underflow in one of the internal calculations. Fixed, \{RD/RN/SR\} refer to fixed-point arithmetic with round-down, round-to-nearest and stochastic rounding, respectively.}
	\label{table:ode-stats-16bit}
\end{table}

\section{Ideas related to \emph{Dither} and how these may be applicable}
\label{sec:dither}

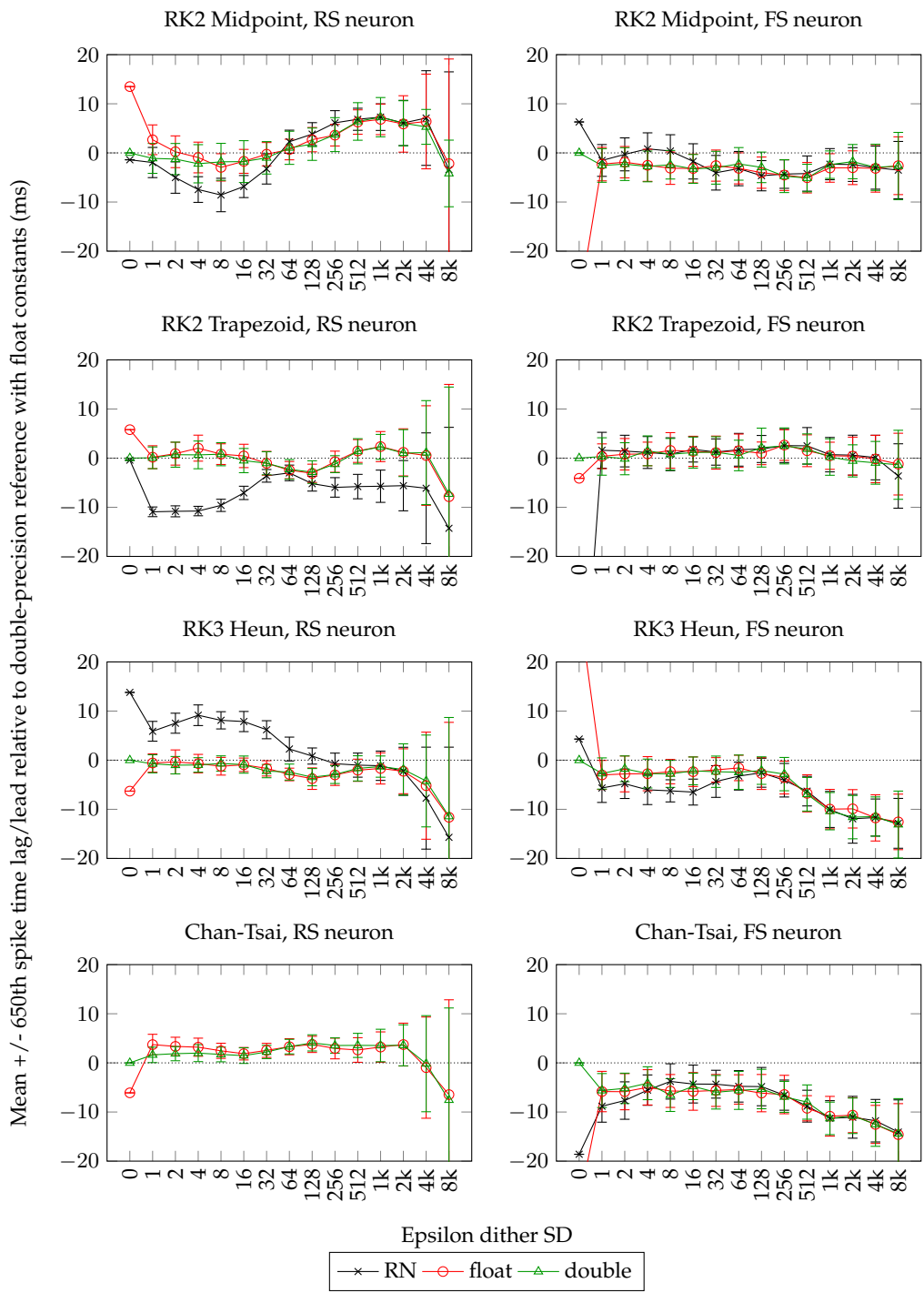
\begin{figure}[ht!]
\begin{center}
    % NOTE: The following avoids the regeneration of the figure and imports the PDF. To regenerate the PDF using the following script
    % run in the terminal 'pdflatex --jobname images/dither-diagram RS_paper_ODE'.
	\beginpgfgraphicnamed{dither-diagram}
\rotatebox[origin=c]{90}{Mean +/- 650th spike time lag/lead relative to double-precision reference with float constants (ms)}
\begin{tabular}{c c}
\begin{tikzpicture}
% let both axes use the same layers
        \begin{axis}[
            xmin=0, xmax=16,
            ymin=-20, ymax=20,
            width=2.7in,
            height=1.75in,
            legend pos=south east,
            legend entries={\text{RN}, %\text{SR}, 
            \text{float}, \text{double} },
            legend to name=legend-dither,
            legend columns=4,
            title={RK2 Midpoint, RS neuron},
            xtick={1, 2, 3, 4, 5, 6, 7, 8, 9, 10, 11, 12, 13, 14, 15},
            xticklabels={0, 1, 2, 4, 8, 16, 32, 64, 128, 256, 512, 1k, 2k, 4k, 8k},
           x tick label style={rotate=90,anchor=east}
        ]
        \addplot+ [color=black, mark=x, error bars/.cd, y dir=both, y explicit]
	coordinates	{								
(	1	,	-1.4	)	+-	(	0	,	0	)
(	2	,	-1.9575	)	+-	(	3.072365454	,	3.072365454	)
(	3	,	-5.0225	)	+-	(	3.190087813	,	3.190087813	)
(	4	,	-7.4575	)	+-	(	2.614641957	,	2.614641957	)
(	5	,	-8.57	)	+-	(	3.403783716	,	3.403783716	)
(	6	,	-6.8475	)	+-	(	2.245107215	,	2.245107215	)
(	7	,	-3.2525	)	+-	(	3.083204085	,	3.083204085	)
(	8	,	2.272499999	)	+-	(	2.353392259	,	2.353392259	)
(	9	,	3.822499999	)	+-	(	2.336169064	,	2.336169064	)
(	10	,	6.12	)	+-	(	2.480818723	,	2.480818723	)
(	11	,	6.844999999	)	+-	(	2.245217138	,	2.245217138	)
(	12	,	7.2825	)	+-	(	2.709043496	,	2.709043496	)
(	13	,	6.09	)	+-	(	4.609510792	,	4.609510792	)
(	14	,	7.1	)	+-	(	9.634366904	,	9.634366904	)
(	15	,	-3.405	)	+-	(	19.90787694	,	19.90787694	)
}	;										    							        
	\addplot+ [color=red, mark=o, error bars/.cd, y dir=both, y explicit]
	coordinates	{								
(	1	,	13.5	)	+-	(	0	,	0	)
(	2	,	2.675	)	+-	(	3.0245788	,	3.0245788	)
(	3	,	0.2325	)	+-	(	3.231764422	,	3.231764422	)
(	4	,	-0.9325	)	+-	(	3.124049599	,	3.124049599	)
(	5	,	-2.945	)	+-	(	2.780974741	,	2.780974741	)
(	6	,	-1.717500001	)	+-	(	2.434041431	,	2.434041431	)
(	7	,	-0.262500001	)	+-	(	2.371160864	,	2.371160864	)
(	8	,	0.6875	)	+-	(	2.095439186	,	2.095439186	)
(	9	,	2.6975	)	+-	(	2.473188277	,	2.473188277	)
(	10	,	3.6125	)	+-	(	2.200371472	,	2.200371472	)
(	11	,	6.285	)	+-	(	2.481268286	,	2.481268286	)
(	12	,	6.832499999	)	+-	(	3.113114319	,	3.113114319	)
(	13	,	5.895	)	+-	(	5.74241752	,	5.74241752	)
(	14	,	6.397499999	)	+-	(	9.637546434	,	9.637546434	)
(	15	,	-2.1875	)	+-	(	21.30606255	,	21.30606255	)
}	;																			
	\addplot+ [color=green!60!black, mark=triangle, solid, error bars/.cd, y dir=both, y explicit]
	coordinates	{								
(	1	,	0	)	+-	(	0	,	0	)
(	2	,	-1.117500001	)	+-	(	3.044329533	,	3.044329533	)
(	3	,	-1.255000001	)	+-	(	3.174816772	,	3.174816772	)
(	4	,	-2.21	)	+-	(	3.88962789	,	3.88962789	)
(	5	,	-1.817500001	)	+-	(	3.765562579	,	3.765562579	)
(	6	,	-1.7675	)	+-	(	4.26039528	,	4.26039528	)
(	7	,	-1	)	+-	(	3.302981364	,	3.302981364	)
(	8	,	1.0375	)	+-	(	3.368922828	,	3.368922828	)
(	9	,	1.7725	)	+-	(	3.292001768	,	3.292001768	)
(	10	,	3.7225	)	+-	(	3.454803419	,	3.454803419	)
(	11	,	6.402499999	)	+-	(	3.824737296	,	3.824737296	)
(	12	,	7.274999999	)	+-	(	3.996516913	,	3.996516913	)
(	13	,	6.045	)	+-	(	4.573813423	,	4.573813423	)
(	14	,	5.3	)	+-	(	3.548405203	,	3.548405203	)
(	15	,	-4.1825	)	+-	(	6.781089845	,	6.781089845	)
}	;									
\addplot+ [color=black, mark=none, densely dotted, domain=0:16]{0}; 
\end{axis}
\end{tikzpicture}				&
\begin{tikzpicture}
% let both axes use the same layers
        \begin{axis}[
            xmin=0, xmax=16,
            ymin=-20, ymax=20,
            width=2.7in,
            height=1.75in,
            title={RK2 Midpoint, FS neuron},
            xtick={1, 2, 3, 4, 5, 6, 7, 8, 9, 10, 11, 12, 13, 14, 15},
            xticklabels={0, 1, 2, 4, 8, 16, 32, 64, 128, 256, 512, 1k, 2k, 4k, 8k},
           x tick label style={rotate=90,anchor=east}
        ]
        \addplot+ [color=black, mark=x, error bars/.cd, y dir=both, y explicit]
	coordinates	{								
(	1	,	6.3	)	+-	(	0	,	0	)
(	2	,	-1.537500001	)	+-	(	3.234881006	,	3.234881006	)
(	3	,	-0.2975	)	+-	(	3.355247516	,	3.355247516	)
(	4	,	0.8225	)	+-	(	3.262961333	,	3.262961333	)
(	5	,	0.382499999	)	+-	(	3.317311817	,	3.317311817	)
(	6	,	-1.725000001	)	+-	(	3.58785344	,	3.58785344	)
(	7	,	-4.04	)	+-	(	3.52331429	,	3.52331429	)
(	8	,	-3.192500001	)	+-	(	3.487515278	,	3.487515278	)
(	9	,	-4.645	)	+-	(	3.074789965	,	3.074789965	)
(	10	,	-4.312500001	)	+-	(	2.897938037	,	2.897938037	)
(	11	,	-4.202500001	)	+-	(	3.5832579	,	3.5832579	)
(	12	,	-2.295000001	)	+-	(	3.170695143	,	3.170695143	)
(	13	,	-2.372500001	)	+-	(	3.437350462	,	3.437350462	)
(	14	,	-3.0025	)	+-	(	4.383915092	,	4.383915092	)
(	15	,	-3.500000001	)	+-	(	5.832358892	,	5.832358892	)
}	;										    								        
	\addplot+ [color=red, mark=o, error bars/.cd, y dir=both, y explicit]
	coordinates	{								
(	1	,	-30.5	)	+-	(	0	,	0	)
(	2	,	-2.280000001	)	+-	(	3.410782299	,	3.410782299	)
(	3	,	-1.895000001	)	+-	(	3.193258122	,	3.193258122	)
(	4	,	-2.522500001	)	+-	(	3.266417115	,	3.266417115	)
(	5	,	-3.110000001	)	+-	(	3.295513655	,	3.295513655	)
(	6	,	-3.17	)	+-	(	2.934166548	,	2.934166548	)
(	7	,	-2.587500001	)	+-	(	3.170471735	,	3.170471735	)
(	8	,	-3.1425	)	+-	(	3.164213962	,	3.164213962	)
(	9	,	-4.017500001	)	+-	(	3.184166557	,	3.184166557	)
(	10	,	-4.5125	)	+-	(	3.133703265	,	3.133703265	)
(	11	,	-5.050000001	)	+-	(	3.099462318	,	3.099462318	)
(	12	,	-3.065	)	+-	(	2.901065233	,	2.901065233	)
(	13	,	-3.015000001	)	+-	(	3.432839147	,	3.432839147	)
(	14	,	-3.150000001	)	+-	(	4.836612503	,	4.836612503	)
(	15	,	-2.605000001	)	+-	(	5.880212363	,	5.880212363	)
}	;																			
	\addplot+ [color=green!60!black, mark=triangle, solid, error bars/.cd, y dir=both, y explicit]
	coordinates	{								
(	1	,	0	)	+-	(	0	,	0	)
(	2	,	-2.56	)	+-	(	3.416611006	,	3.416611006	)
(	3	,	-2.275	)	+-	(	3.339871792	,	3.339871792	)
(	4	,	-2.752500001	)	+-	(	3.127298129	,	3.127298129	)
(	5	,	-2.410000001	)	+-	(	2.904620545	,	2.904620545	)
(	6	,	-3.260000001	)	+-	(	2.935092712	,	2.935092712	)
(	7	,	-3.0175	)	+-	(	3.341033733	,	3.341033733	)
(	8	,	-2.232500001	)	+-	(	3.319050489	,	3.319050489	)
(	9	,	-2.94	)	+-	(	3.107955053	,	3.107955053	)
(	10	,	-4.7325	)	+-	(	3.339536855	,	3.339536855	)
(	11	,	-5.12	)	+-	(	2.803861805	,	2.803861805	)
(	12	,	-2.392500001	)	+-	(	2.799394394	,	2.799394394	)
(	13	,	-1.7725	)	+-	(	3.482924463	,	3.482924463	)
(	14	,	-2.932500001	)	+-	(	4.702175573	,	4.702175573	)
(	15	,	-2.6775	)	+-	(	6.84991063	,	6.84991063	)
}	;									
\addplot+ [color=black, mark=none, densely dotted, domain=0:16]{0}; 
\end{axis}
\end{tikzpicture} \\
\begin{tikzpicture}
% let both axes use the same layers
        \begin{axis}[
            xmin=0, xmax=16,
            ymin=-20, ymax=20,
            width=2.7in,
            height=1.75in,
            title={RK2 Trapezoid, RS neuron},
            xtick={1, 2, 3, 4, 5, 6, 7, 8, 9, 10, 11, 12, 13, 14, 15},
            xticklabels={0, 1, 2, 4, 8, 16, 32, 64, 128, 256, 512, 1k, 2k, 4k, 8k},
           x tick label style={rotate=90,anchor=east}
        ]
        \addplot+ [color=black, mark=x, error bars/.cd, y dir=both, y explicit]
	coordinates	{								
(	1	,	-0.4	)	+-	(	0	,	0	)
(	2	,	-10.92	)	+-	(	0.983244237	,	0.983244237	)
(	3	,	-10.82	)	+-	(	1.116128841	,	1.116128841	)
(	4	,	-10.765	)	+-	(	0.960381868	,	0.960381868	)
(	5	,	-9.607500001	)	+-	(	1.241047427	,	1.241047427	)
(	6	,	-7.070000001	)	+-	(	1.354990774	,	1.354990774	)
(	7	,	-3.5975	)	+-	(	1.288506887	,	1.288506887	)
(	8	,	-3.0025	)	+-	(	1.589305202	,	1.589305202	)
(	9	,	-5.1925	)	+-	(	1.511390512	,	1.511390512	)
(	10	,	-5.952500001	)	+-	(	1.985845747	,	1.985845747	)
(	11	,	-5.785	)	+-	(	2.485295215	,	2.485295215	)
(	12	,	-5.715000001	)	+-	(	3.281771722	,	3.281771722	)
(	13	,	-5.6	)	+-	(	5.119745586	,	5.119745586	)
(	14	,	-6.1225	)	+-	(	11.26727705	,	11.26727705	)
(	15	,	-14.2775	)	+-	(	20.55459881	,	20.55459881	)
}	;																										
	        \addplot+ [color=red, mark=o, error bars/.cd, y dir=both, y explicit]
	coordinates	{								
(	1	,	5.8	)	+-	(	0	,	0	)
(	2	,	0.1925	)	+-	(	2.333302655	,	2.333302655	)
(	3	,	0.914999999	)	+-	(	2.329691296	,	2.329691296	)
(	4	,	2.0375	)	+-	(	2.645430264	,	2.645430264	)
(	5	,	0.817499999	)	+-	(	2.105534526	,	2.105534526	)
(	6	,	0.4525	)	+-	(	2.375946505	,	2.375946505	)
(	7	,	-1.042500001	)	+-	(	2.424224697	,	2.424224697	)
(	8	,	-2.435000001	)	+-	(	1.75609561	,	1.75609561	)
(	9	,	-3.087500001	)	+-	(	1.858202536	,	1.858202536	)
(	10	,	-0.722500001	)	+-	(	2.15269509	,	2.15269509	)
(	11	,	1.457499999	)	+-	(	2.323225818	,	2.323225818	)
(	12	,	2.354999999	)	+-	(	3.06133455	,	3.06133455	)
(	13	,	1.18	)	+-	(	4.796376837	,	4.796376837	)
(	14	,	0.5325	)	+-	(	10.13646344	,	10.13646344	)
(	15	,	-7.85	)	+-	(	22.86453978	,	22.86453978	)
}	;																			
       \addplot+ [color=green!60!black, mark=triangle, solid, error bars/.cd, y dir=both, y explicit]
	coordinates	{								
(	1	,	0	)	+-	(	0	,	0	)
(	2	,	0.075	)	+-	(	2.210087595	,	2.210087595	)
(	3	,	0.7075	)	+-	(	2.54773781	,	2.54773781	)
(	4	,	0.665	)	+-	(	2.822124131	,	2.822124131	)
(	5	,	0.805	)	+-	(	2.353497125	,	2.353497125	)
(	6	,	-0.495000001	)	+-	(	2.464096027	,	2.464096027	)
(	7	,	-1.037500001	)	+-	(	2.339481369	,	2.339481369	)
(	8	,	-2.285	)	+-	(	2.000583248	,	2.000583248	)
(	9	,	-2.8375	)	+-	(	2.302806259	,	2.302806259	)
(	10	,	-1.225000001	)	+-	(	1.725339059	,	1.725339059	)
(	11	,	1.3925	)	+-	(	2.569883515	,	2.569883515	)
(	12	,	2.2975	)	+-	(	2.53736058	,	2.53736058	)
(	13	,	1.0275	)	+-	(	4.773752354	,	4.773752354	)
(	14	,	1.127499999	)	+-	(	10.58368332	,	10.58368332	)
(	15	,	-7.287500001	)	+-	(	21.76160094	,	21.76160094	)
}	;									      
\addplot+ [color=black, mark=none, densely dotted, domain=0:16]{0}; 
        \end{axis}
\end{tikzpicture}
&
\begin{tikzpicture}
% let both axes use the same layers
        \begin{axis}[
            xmin=0, xmax=16,
            ymin=-20, ymax=20,
            width=2.7in,
            height=1.75in,
            title={RK2 Trapezoid, FS neuron},
            xtick={1, 2, 3, 4, 5, 6, 7, 8, 9, 10, 11, 12, 13, 14, 15},
            xticklabels={0, 1, 2, 4, 8, 16, 32, 64, 128, 256, 512, 1k, 2k, 4k, 8k},
           x tick label style={rotate=90,anchor=east}
        ]
        \addplot+ [color=black, mark=x, error bars/.cd, y dir=both, y explicit]
	coordinates	{								
(	1	,	-61.8	)	+-	(	0	,	0	)
(	2	,	1.599999999	)	+-	(	3.660110689	,	3.660110689	)
(	3	,	1.4425	)	+-	(	3.21964224	,	3.21964224	)
(	4	,	1.152499999	)	+-	(	3.249614375	,	3.249614375	)
(	5	,	0.804999999	)	+-	(	3.343033322	,	3.343033322	)
(	6	,	1.7875	)	+-	(	2.579424254	,	2.579424254	)
(	7	,	1.247499999	)	+-	(	2.673131986	,	2.673131986	)
(	8	,	1.71	)	+-	(	3.330881149	,	3.330881149	)
(	9	,	1.865	)	+-	(	2.753044934	,	2.753044934	)
(	10	,	2.549999999	)	+-	(	3.449860643	,	3.449860643	)
(	11	,	2.509999999	)	+-	(	3.71889148	,	3.71889148	)
(	12	,	0.71	)	+-	(	3.502877937	,	3.502877937	)
(	13	,	0.612499999	)	+-	(	4.060831511	,	4.060831511	)
(	14	,	0.124999999	)	+-	(	4.539216582	,	4.539216582	)
(	15	,	-3.63	)	+-	(	6.554239184	,	6.554239184	)
}	;																										
	        \addplot+ [color=red, mark=o, error bars/.cd, y dir=both, y explicit]
	coordinates	{								
(	1	,	-4.1	)	+-	(	0	,	0	)
(	2	,	0.379999999	)	+-	(	2.528118789	,	2.528118789	)
(	3	,	0.797499999	)	+-	(	3.200840234	,	3.200840234	)
(	4	,	0.8575	)	+-	(	2.400522112	,	2.400522112	)
(	5	,	1.577499999	)	+-	(	3.620347589	,	3.620347589	)
(	6	,	1.39	)	+-	(	3.070438037	,	3.070438037	)
(	7	,	1.129999999	)	+-	(	3.365450402	,	3.365450402	)
(	8	,	1.515	)	+-	(	3.387196859	,	3.387196859	)
(	9	,	0.98	)	+-	(	2.304199065	,	2.304199065	)
(	10	,	2.665	)	+-	(	3.116338252	,	3.116338252	)
(	11	,	1.509999999	)	+-	(	3.240117123	,	3.240117123	)
(	12	,	0.484999999	)	+-	(	2.761507392	,	2.761507392	)
(	13	,	0.41	)	+-	(	3.870652234	,	3.870652234	)
(	14	,	-0.165000001	)	+-	(	4.817412328	,	4.817412328	)
(	15	,	-1.212500001	)	+-	(	6.283400496	,	6.283400496	)
}	;																			
       \addplot+ [color=green!60!black, mark=triangle, solid, error bars/.cd, y dir=both, y explicit]
	coordinates	{								
(	1	,	0	)	+-	(	0	,	0	)
(	2	,	0.339999999	)	+-	(	3.809690747	,	3.809690747	)
(	3	,	-0.120000001	)	+-	(	3.259085918	,	3.259085918	)
(	4	,	1.49	)	+-	(	3.045197985	,	3.045197985	)
(	5	,	0.785	)	+-	(	3.222321984	,	3.222321984	)
(	6	,	1.147499999	)	+-	(	3.171021634	,	3.171021634	)
(	7	,	1.212499999	)	+-	(	3.152141111	,	3.152141111	)
(	8	,	0.5475	)	+-	(	3.107475477	,	3.107475477	)
(	9	,	2.149999999	)	+-	(	3.960963363	,	3.960963363	)
(	10	,	2.4925	)	+-	(	3.663568912	,	3.663568912	)
(	11	,	1.8675	)	+-	(	3.072323725	,	3.072323725	)
(	12	,	0.2275	)	+-	(	3.710483243	,	3.710483243	)
(	13	,	-0.555000001	)	+-	(	3.270532772	,	3.270532772	)
(	14	,	-0.95	)	+-	(	4.35024314	,	4.35024314	)
(	15	,	-1.3525	)	+-	(	7.025192305	,	7.025192305	)
}	;									      
\addplot+ [color=black, mark=none, densely dotted, domain=0:16]{0}; 
        \end{axis}
\end{tikzpicture} \\
\begin{tikzpicture}
% let both axes use the same layers
        \begin{axis}[
            xmin=0, xmax=16,
            ymin=-20, ymax=20,
            width=2.7in,
            height=1.75in,
            title={RK3 Heun, RS neuron},
            xtick={1, 2, 3, 4, 5, 6, 7, 8, 9, 10, 11, 12, 13, 14, 15},
            xticklabels={0, 1, 2, 4, 8, 16, 32, 64, 128, 256, 512, 1k, 2k, 4k, 8k},
           x tick label style={rotate=90,anchor=east}
        ]
        \addplot+ [color=black, mark=x, error bars/.cd, y dir=both, y explicit]
	coordinates	{								
(	1	,	13.8	)	+-	(	0	,	0	)
(	2	,	5.885	)	+-	(	2.01526864	,	2.01526864	)
(	3	,	7.527499999	)	+-	(	2.03709037	,	2.03709037	)
(	4	,	9.149999999	)	+-	(	2.101769706	,	2.101769706	)
(	5	,	8.1175	)	+-	(	1.753076051	,	1.753076051	)
(	6	,	7.867499999	)	+-	(	2.051314462	,	2.051314462	)
(	7	,	6.199999999	)	+-	(	1.835127327	,	1.835127327	)
(	8	,	2.272499999	)	+-	(	2.428567954	,	2.428567954	)
(	9	,	0.847499999	)	+-	(	1.643009668	,	1.643009668	)
(	10	,	-0.712500001	)	+-	(	2.154207272	,	2.154207272	)
(	11	,	-1.005000001	)	+-	(	2.4515772	,	2.4515772	)
(	12	,	-1.1325	)	+-	(	2.980654505	,	2.980654505	)
(	13	,	-2.262500001	)	+-	(	4.890843747	,	4.890843747	)
(	14	,	-7.727500001	)	+-	(	10.37951601	,	10.37951601	)
(	15	,	-15.705	)	+-	(	18.35716796	,	18.35716796	)
}	;																							
	        \addplot+ [color=red, mark=o, error bars/.cd, y dir=both, y explicit]
	coordinates	{								
(	1	,	-6.3	)	+-	(	0	,	0	)
(	2	,	-0.572500001	)	+-	(	1.856794522	,	1.856794522	)
(	3	,	-0.3525	)	+-	(	2.440364359	,	2.440364359	)
(	4	,	-0.6975	)	+-	(	1.879373531	,	1.879373531	)
(	5	,	-1.235000001	)	+-	(	1.799650963	,	1.799650963	)
(	6	,	-0.935000001	)	+-	(	1.369549804	,	1.369549804	)
(	7	,	-1.665	)	+-	(	1.557537504	,	1.557537504	)
(	8	,	-2.902500001	)	+-	(	1.280121688	,	1.280121688	)
(	9	,	-3.7575	)	+-	(	2.189237954	,	2.189237954	)
(	10	,	-2.995	)	+-	(	2.14595027	,	2.14595027	)
(	11	,	-2.045	)	+-	(	2.258879913	,	2.258879913	)
(	12	,	-1.675	)	+-	(	3.174558317	,	3.174558317	)
(	13	,	-2.255	)	+-	(	4.57585357	,	4.57585357	)
(	14	,	-5.19	)	+-	(	10.89979534	,	10.89979534	)
(	15	,	-11.6225	)	+-	(	19.32760066	,	19.32760066	)
}	;									
										
       \addplot+ [color=green!60!black, mark=triangle, solid, error bars/.cd, y dir=both, y explicit]
	coordinates	{								
(	1	,	0	)	+-	(	0	,	0	)
(	2	,	-0.735	)	+-	(	1.839112104	,	1.839112104	)
(	3	,	-1.0175	)	+-	(	1.769671489	,	1.769671489	)
(	4	,	-0.965000001	)	+-	(	1.475535539	,	1.475535539	)
(	5	,	-0.6725	)	+-	(	1.555797676	,	1.555797676	)
(	6	,	-0.895	)	+-	(	1.745169156	,	1.745169156	)
(	7	,	-2.22	)	+-	(	1.350251638	,	1.350251638	)
(	8	,	-2.417500001	)	+-	(	1.628100592	,	1.628100592	)
(	9	,	-3.457500001	)	+-	(	1.761538881	,	1.761538881	)
(	10	,	-2.92	)	+-	(	1.825348579	,	1.825348579	)
(	11	,	-1.665000001	)	+-	(	2.566704961	,	2.566704961	)
(	12	,	-1.305000001	)	+-	(	2.16142759	,	2.16142759	)
(	13	,	-1.8875	)	+-	(	5.214830306	,	5.214830306	)
(	14	,	-4.23	)	+-	(	9.347650191	,	9.347650191	)
(	15	,	-11.435	)	+-	(	20.13059351	,	20.13059351	)
}	;									      
	\addplot+ [color=black, mark=none, densely dotted, domain=0:16]{0}; 
        \end{axis}
\end{tikzpicture}			&
\begin{tikzpicture}
% let both axes use the same layers
        \begin{axis}[
            xmin=0, xmax=16,
            ymin=-20, ymax=20,
            width=2.7in,
            height=1.75in,
            title={RK3 Heun, FS neuron},
            xtick={1, 2, 3, 4, 5, 6, 7, 8, 9, 10, 11, 12, 13, 14, 15},
            xticklabels={0, 1, 2, 4, 8, 16, 32, 64, 128, 256, 512, 1k, 2k, 4k, 8k},
           x tick label style={rotate=90,anchor=east}
        ]
        \addplot+ [color=black, mark=x, error bars/.cd, y dir=both, y explicit]
	coordinates	{								
(	1	,	4.3	)	+-	(	0	,	0	)
(	2	,	-5.6575	)	+-	(	2.941296373	,	2.941296373	)
(	3	,	-4.7975	)	+-	(	2.998759358	,	2.998759358	)
(	4	,	-6.030000001	)	+-	(	3.005823407	,	3.005823407	)
(	5	,	-6.24	)	+-	(	2.25261102	,	2.25261102	)
(	6	,	-6.4925	)	+-	(	2.639870577	,	2.639870577	)
(	7	,	-4.340000001	)	+-	(	3.225125718	,	3.225125718	)
(	8	,	-3.235000001	)	+-	(	2.841501047	,	2.841501047	)
(	9	,	-2.555	)	+-	(	2.947571795	,	2.947571795	)
(	10	,	-4.075	)	+-	(	3.415931753	,	3.415931753	)
(	11	,	-6.3025	)	+-	(	3.007127217	,	3.007127217	)
(	12	,	-10.03	)	+-	(	3.67166559	,	3.67166559	)
(	13	,	-11.9275	)	+-	(	4.965728056	,	4.965728056	)
(	14	,	-11.675	)	+-	(	3.774119017	,	3.774119017	)
(	15	,	-12.855	)	+-	(	5.077145877	,	5.077145877	)
}	;																																						
	        \addplot+ [color=red, mark=o, error bars/.cd, y dir=both, y explicit]
	coordinates	{								
(	1	,	31.2	)	+-	(	0	,	0	)
(	2	,	-3.002500001	)	+-	(	2.958385517	,	2.958385517	)
(	3	,	-2.795	)	+-	(	3.649794514	,	3.649794514	)
(	4	,	-2.752500001	)	+-	(	3.47570229	,	3.47570229	)
(	5	,	-2.3325	)	+-	(	2.481251493	,	2.481251493	)
(	6	,	-2.345	)	+-	(	3.011937786	,	3.011937786	)
(	7	,	-1.9725	)	+-	(	2.490596417	,	2.490596417	)
(	8	,	-1.577500001	)	+-	(	2.589375481	,	2.589375481	)
(	9	,	-2.722500001	)	+-	(	3.235578456	,	3.235578456	)
(	10	,	-3.5325	)	+-	(	3.342146362	,	3.342146362	)
(	11	,	-6.745000001	)	+-	(	3.768761613	,	3.768761613	)
(	12	,	-9.962500001	)	+-	(	4.01833538	,	4.01833538	)
(	13	,	-9.9025	)	+-	(	3.907815147	,	3.907815147	)
(	14	,	-11.7475	)	+-	(	4.701553889	,	4.701553889	)
(	15	,	-12.5625	)	+-	(	5.65325808	,	5.65325808	)
}	;																	
										
       \addplot+ [color=green!60!black, mark=triangle, solid, error bars/.cd, y dir=both, y explicit]
	coordinates	{								
(	1	,	0	)	+-	(	0	,	0	)
(	2	,	-2.71	)	+-	(	3.178759635	,	3.178759635	)
(	3	,	-1.752500001	)	+-	(	2.618179784	,	2.618179784	)
(	4	,	-2.7175	)	+-	(	2.942342299	,	2.942342299	)
(	5	,	-2.787500001	)	+-	(	2.750215609	,	2.750215609	)
(	6	,	-2.175	)	+-	(	2.829763965	,	2.829763965	)
(	7	,	-2.350000001	)	+-	(	3.186992795	,	3.186992795	)
(	8	,	-2.445	)	+-	(	3.493873759	,	3.493873759	)
(	9	,	-2.117500001	)	+-	(	2.801912121	,	2.801912121	)
(	10	,	-2.832500001	)	+-	(	3.389757384	,	3.389757384	)
(	11	,	-6.937500001	)	+-	(	3.423050633	,	3.423050633	)
(	12	,	-10.3775	)	+-	(	3.796049734	,	3.796049734	)
(	13	,	-11.5925	)	+-	(	4.409627433	,	4.409627433	)
(	14	,	-11.47	)	+-	(	3.990385881	,	3.990385881	)
(	15	,	-13.1	)	+-	(	6.813561138	,	6.813561138	)
}	;																      
	\addplot+ [color=black, mark=none, densely dotted, domain=0:16]{0}; 
        \end{axis}
\end{tikzpicture} \\
\begin{tikzpicture}
% let both axes use the same layers
        \begin{axis}[
            xmin=0, xmax=16,
            ymin=-20, ymax=20,
            width=2.7in,
            height=1.75in,
            title={Chan-Tsai, RS neuron},
            xtick={1, 2, 3, 4, 5, 6, 7, 8, 9, 10, 11, 12, 13, 14, 15},
            xticklabels={0, 1, 2, 4, 8, 16, 32, 64, 128, 256, 512, 1k, 2k, 4k, 8k},
           x tick label style={rotate=90,anchor=east}
        ]
        \addplot+ [color=black, mark=x, error bars/.cd, y dir=both, y explicit]
	coordinates	{								
(	1	,	-158.5	)	+-	(	0	,	0	)
(	2	,	-104.65	)	+-	(	2.458267056	,	2.458267056	)
(	3	,	-105.0125	)	+-	(	3.463375224	,	3.463375224	)
(	4	,	-103.96	)	+-	(	2.967412755	,	2.967412755	)
(	5	,	-104.0875	)	+-	(	2.997365296	,	2.997365296	)
(	6	,	-102.015	)	+-	(	2.611959672	,	2.611959672	)
(	7	,	-102.9275	)	+-	(	2.632147168	,	2.632147168	)
(	8	,	-103.9875	)	+-	(	2.139037346	,	2.139037346	)
(	9	,	-100.7375	)	+-	(	2.589989357	,	2.589989357	)
(	10	,	-95.935	)	+-	(	2.211108319	,	2.211108319	)
(	11	,	-93.3575	)	+-	(	2.447593009	,	2.447593009	)
(	12	,	-94.76	)	+-	(	2.975782594	,	2.975782594	)
(	13	,	-96.83	)	+-	(	4.488001096	,	4.488001096	)
(	14	,	-94.3275	)	+-	(	10.76959127	,	10.76959127	)
(	15	,	-105.57	)	+-	(	17.92084018	,	17.92084018	)
}	;																		       
 \addplot+ [color=red, mark=o, error bars/.cd, y dir=both, y explicit]
	coordinates	{								
(	1	,	-6.1	)	+-	(	0	,	0	)
(	2	,	3.717499999	)	+-	(	2.103097532	,	2.103097532	)
(	3	,	3.352499999	)	+-	(	1.867260093	,	1.867260093	)
(	4	,	3.1925	)	+-	(	1.842154187	,	1.842154187	)
(	5	,	2.4575	)	+-	(	1.542706163	,	1.542706163	)
(	6	,	1.864999999	)	+-	(	1.266238119	,	1.266238119	)
(	7	,	2.547499999	)	+-	(	1.43222751	,	1.43222751	)
(	8	,	3.2875	)	+-	(	1.63897834	,	1.63897834	)
(	9	,	3.769999999	)	+-	(	1.649739684	,	1.649739684	)
(	10	,	2.929999999	)	+-	(	2.108006714	,	2.108006714	)
(	11	,	2.624999999	)	+-	(	2.501973579	,	2.501973579	)
(	12	,	3.219999999	)	+-	(	3.089535688	,	3.089535688	)
(	13	,	3.7275	)	+-	(	4.359427597	,	4.359427597	)
(	14	,	-0.965	)	+-	(	10.30059869	,	10.30059869	)
(	15	,	-6.46	)	+-	(	19.30286016	,	19.30286016	)
}	;									
										
       \addplot+ [color=green!60!black, mark=triangle, solid, error bars/.cd, y dir=both, y explicit]
	coordinates	{								
(	1	,	0	)	+-	(	0	,	0	)
(	2	,	1.6325	)	+-	(	1.578442513	,	1.578442513	)
(	3	,	1.872499999	)	+-	(	1.438836335	,	1.438836335	)
(	4	,	1.9875	)	+-	(	1.729856596	,	1.729856596	)
(	5	,	1.719999999	)	+-	(	1.500632345	,	1.500632345	)
(	6	,	1.4775	)	+-	(	1.587852524	,	1.587852524	)
(	7	,	2.217499999	)	+-	(	1.348672186	,	1.348672186	)
(	8	,	3.3075	)	+-	(	1.483652371	,	1.483652371	)
(	9	,	4.07	)	+-	(	1.644290822	,	1.644290822	)
(	10	,	3.534999999	)	+-	(	1.557537504	,	1.557537504	)
(	11	,	3.562499999	)	+-	(	2.454267607	,	2.454267607	)
(	12	,	3.539999999	)	+-	(	3.306078318	,	3.306078318	)
(	13	,	3.574999999	)	+-	(	4.177028142	,	4.177028142	)
(	14	,	-0.165	)	+-	(	9.7796374	,	9.7796374	)
(	15	,	-7.5625	)	+-	(	18.73990241	,	18.73990241	)
}	;									      \addplot+ [color=black, mark=none, densely dotted, domain=0:16]{0}; 
        \end{axis}
\end{tikzpicture}
&
\begin{tikzpicture}
% let both axes use the same layers
        \begin{axis}[
            xmin=0, xmax=16,
            ymin=-20, ymax=20,
            width=2.7in,
            height=1.75in,
            title={Chan-Tsai, FS neuron},
            xtick={1, 2, 3, 4, 5, 6, 7, 8, 9, 10, 11, 12, 13, 14, 15},
            xticklabels={0, 1, 2, 4, 8, 16, 32, 64, 128, 256, 512, 1k, 2k, 4k, 8k},
           x tick label style={rotate=90,anchor=east}
        ]
        \addplot+ [color=black, mark=x, error bars/.cd, y dir=both, y explicit]
	coordinates	{								
(	1	,	-18.6	)	+-	(	0	,	0	)
(	2	,	-8.815	)	+-	(	3.255098564	,	3.255098564	)
(	3	,	-7.6675	)	+-	(	3.810335976	,	3.810335976	)
(	4	,	-5.5425	)	+-	(	3.088181566	,	3.088181566	)
(	5	,	-3.775	)	+-	(	3.592781223	,	3.592781223	)
(	6	,	-4.302500001	)	+-	(	3.858223817	,	3.858223817	)
(	7	,	-4.3375	)	+-	(	2.834647131	,	2.834647131	)
(	8	,	-4.775	)	+-	(	3.20326356	,	3.20326356	)
(	9	,	-4.83	)	+-	(	3.897810036	,	3.897810036	)
(	10	,	-6.567500001	)	+-	(	3.069735435	,	3.069735435	)
(	11	,	-8.805	)	+-	(	3.239416693	,	3.239416693	)
(	12	,	-11.2875	)	+-	(	3.599158555	,	3.599158555	)
(	13	,	-11.0725	)	+-	(	4.256999515	,	4.256999515	)
(	14	,	-11.77	)	+-	(	4.32833242	,	4.32833242	)
(	15	,	-14.0875	)	+-	(	6.554082208	,	6.554082208	)
}	;																	
\addplot+ [color=red, mark=o, error bars/.cd, y dir=both, y explicit]
	coordinates	{								
(	1	,	-28.4	)	+-	(	0	,	0	)
(	2	,	-5.832500001	)	+-	(	4.107990181	,	4.107990181	)
(	3	,	-5.865000001	)	+-	(	3.679015336	,	3.679015336	)
(	4	,	-4.975	)	+-	(	3.593494834	,	3.593494834	)
(	5	,	-5.717500001	)	+-	(	3.350460167	,	3.350460167	)
(	6	,	-5.805000001	)	+-	(	3.832850582	,	3.832850582	)
(	7	,	-5.615000001	)	+-	(	3.24555317	,	3.24555317	)
(	8	,	-5.4075	)	+-	(	2.986155018	,	2.986155018	)
(	9	,	-6.155000001	)	+-	(	3.795608123	,	3.795608123	)
(	10	,	-6.440000001	)	+-	(	3.911836082	,	3.911836082	)
(	11	,	-9.235	)	+-	(	2.563406187	,	2.563406187	)
(	12	,	-10.8575	)	+-	(	4.037318702	,	4.037318702	)
(	13	,	-10.6475	)	+-	(	3.553834151	,	3.553834151	)
(	14	,	-12.525	)	+-	(	3.877003204	,	3.877003204	)
(	15	,	-14.6025	)	+-	(	6.30602022	,	6.30602022	)
}	;																			
\addplot+ [color=green!60!black, mark=triangle, solid, error bars/.cd, y dir=both, y explicit]
	coordinates	{								
(	1	,	0	)	+-	(	0	,	0	)
(	2	,	-5.625000001	)	+-	(	3.453779052	,	3.453779052	)
(	3	,	-5.185	)	+-	(	3.093462917	,	3.093462917	)
(	4	,	-4.14	)	+-	(	3.348692818	,	3.348692818	)
(	5	,	-6.547500001	)	+-	(	3.342038952	,	3.342038952	)
(	6	,	-4.785	)	+-	(	2.691158362	,	2.691158362	)
(	7	,	-5.975000001	)	+-	(	3.417732791	,	3.417732791	)
(	8	,	-5.600000001	)	+-	(	3.870599238	,	3.870599238	)
(	9	,	-5.322500001	)	+-	(	4.012319969	,	4.012319969	)
(	10	,	-6.915	)	+-	(	3.148996584	,	3.148996584	)
(	11	,	-7.9775	)	+-	(	3.488772284	,	3.488772284	)
(	12	,	-11.295	)	+-	(	3.30942322	,	3.30942322	)
(	13	,	-10.7675	)	+-	(	3.610553441	,	3.610553441	)
(	14	,	-12.51	)	+-	(	4.47275513	,	4.47275513	)
(	15	,	-14.4025	)	+-	(	7.125054655	,	7.125054655	)
}	;
\addplot+ [color=black, mark=none, densely dotted, domain=0:16]{0}; 
\end{axis}
\end{tikzpicture}
\\
\multicolumn{2}{c}{Epsilon dither SD} \\
\multicolumn{2}{c}{\ref*{legend-dither}}
\end{tabular}
\endpgfgraphicnamed
\end{center}
\caption{Effects of dither on DC input in the RS and FS neurons with 4 different solvers (single-precision floating-point constants).}
\label{fig:rs-dither}
\end{figure}

We mentioned in the introduction the idea of \emph{dither} first applied in DSP for audio, image and video signals. The idea in essence is that as the input signal gets small enough to be close to the LSB of the digital system, some added noise helps to improve effective resolution of the system below the LSB. This gain is at the expense of a slight increase in broadband noise but, importantly for DSP applications, this noise is now effectively (or in certain cases \emph{exactly}) uncorrelated with respect to the signal.  This is in contrast to quantisation errors which are highly correlated with the signal, and are considered to be objectionable and easy to identify in both audio and visual applications. 

In our context, psychological considerations are difficult to justify.  However, there are many other parallels between audio processing in particular and the solution of ODEs. Both are taking an input signal and using limited precision digital arithmetic operations in order to produce an output over a time sequence where any errors will be time-averaged. The ability of dither effectively to increase the resolution of the digital arithmetic operations was seen as an opportunity to discover if the benefits demonstrated earlier by SR on every multiply could be at least partially replicated by adding an appropriate quantity and distribution of random noise at the input to the ODE. In our cases this is a DC current synaptic input, though any benefits would be generally applicable to the time-varying and conductance inputs that are also common in neural simulations.  

A valuable text \cite{cc96} contains much useful material about the precision and stability of finite precision computations. In particular, chapter 8 describes the PRECISE toolbox which appears to use a similar idea to our application of dither. However, there are arguably two differences: (1) it is primarily a software technique for producing a number of sophisticated sensitivity analyses rather than a method for efficiently improving the accuracy of results and (2) averaging over many iterations in the solution of an ODE - and thereby canceling out individual errors over time - is likely to be an important ingredient in the results described below.

Another analysis that helped lead to this idea was a consideration of the \emph{backward error} of an algorithm \cite[\S1.5]{Higham:2002:ASN}. The most obvious consideration of error is that of \emph{forward error} i.e. for a given input $x$ and algorithm $f(x)$ there is an output $\hat{f}(x)$ which represents a realistic and imperfect algorithm that uses finite-precision arithmetic.  The difference between $f(x)$ and $\hat{f}(x)$ is the forward error. However, it has been shown that in numerical work the backward error is often a more useful measure. Imagine that you want to make $\hat{f}(x)$ match the known ideal $f(x)$ and all you are allowed to do is perturb $x$ to $\hat{x}$ so that theoretically $\hat{f}(\hat{x}) \equiv f(x)$.  The difference between $x$ and $\hat{x}$ now describes the backward error and in many important real world calculations bounds or distributions can be found for this value. It is important to consider how large this value is relative to the precision of your arithmetic type.  If backward error is no greater than precision then the argument is that the error in a result from the algorithm is defined as much as, or more by, the precision of the input than it is by any imperfection in the algorithm itself.

One can now relate this to the dither idea explored above. Now $x$ can be seen as the input current to the neuron, $f(x)$ as the ODE solution method which updates the state variable(s) at every timestep, and dither which adds random noise of some form and size to $x$ becomes a probabilistic embodiment of the backward error of the algorithm.  The hypothesis is that an appropriate dither signal introduced at the input can be found which is (1) no greater than the backward error, and (2) enough to ensure that stochasticity is induced in the arithmetic operations within the algorithm, thereby achieving indirectly a form of SR. 

To identify appropriate values of dither, an empirical approach was used which adds Gaussian noise at the input scaled so that the SD ranges from 1 LSB for the s16.15 type (i.e. 3.05e-5) up to 10,000 LSBs. The DC input is 4.775 nA so that a dither value of 156 would give the input a Coefficient of Variation of 0.1\%. 40 repeats of the eight neuron and solver combinations were then simulated. Unlike the earlier plots, 23-bit float constants were used in all cases which ensures that any offsets caused by differences in precision of the constants are corrected for and so any differences are now entirely due to the arithmetic differences caused by operations and how they are modified by the dither input.

Figure~\ref{fig:rs-dither} 
%and \ref{fig:fs-dither} 
shows the results with the epsilon multiplier on the X axis with 0 meaning no dither and so equivalent to the reference results (except for the difference in constants for all except the \emph{float} result). The Y axis shows the mean difference in the 650th spike time from the \emph{double} reference result over 40 runs with error bars showing the variation in this difference as a standard deviation. We have chosen a Y axis scaling small enough to show subtle differences between the arithmetics, so in some cases very poor results are off the plot and will be described in the text.

The key points from these plots are the following:

\begin{itemize}
   \item In all cases, \emph{float} results are significantly improved by even the smallest amount of dither. This is particularly noticeable on the FS neuron but at this point there is no obvious explanation for this.
   \item In many cases, the smallest amount of dither also produces a significant improvement in the RN results. This is not the case for the RK2 solvers used with the RS neuron which perform well with 0 dither and small amounts of dither initially worsen the performance. The Chan-Tsai solver with the RS neuron has very high errors with RN and any amount of dither.
	\item The SD of the results is relatively consistent with large amounts of dither until a threshold, which is at approximately 1k for the RS neuron and approximately 4k for the FS neuron. The mean values drift away from the best results at approximately the same dither values. In some cases increasing input noise appears actually to reduce the spike lag SDs until they increase again at this threshold, e.g. both RK2 solvers with the RS neuron.  This would seem counter-intuitive unless something other than random variation is relevant.
	\item Although not shown on the plots for clarity, SR tracks \emph{double} better than RN and this is more apparent on the RS neuron. It is interesting to note in the full results how \emph{float}, \emph{double} and SR  follow similar trends over dither value.
	\item There is an interesting apparent agreement between the results at approx 32-128 dither with all arithmetics (except RN in some cases) converging on the same result and perhaps the SD of the results also reducing slightly. This is more apparent on the RS neuron. There are a number of possible mechanisms for this effect - if it actually exists and is not just sampling variation - and it would be useful to understand them in terms of optimising precision by choosing an appropriate dither value. There is an intriguing (but speculative) possibility that results at these dither values are converging on a result with less algorithmic error than the \emph{double} arithmetic reference which itself has finite precision. Further research is needed to confirm or refute these hypotheses.
    \item We have achieved similar results with stochastic rounding of all the constants involved in the ODE. The method was to choose one of the two versions of each constant on each ODE integration step stochastically, based on a more precise version of that constant. The results were approximately equivalent in 2nd order solvers.
\end{itemize}

Dither is in most cases going to be significantly simpler and cheaper than SR to implement, and for some applications this will be an important consideration. Runtime will be approximately $N$ times faster if there are $N$ operations within one iteration of the solver. For example, one of the ESR RK2 solvers investigated here uses 14 multiply operations. However, results so far are not as consistent or robust as SR. It appears to work better on the RK2 solvers. Our current hypothesis is that SR always induces the ideal uniform stochasticity for each multiply whereas in the case of dither the higher the order of the solver algorithm the more complex the processing of the input and hence the more distorted the Gaussian noise will become as it is increasingly transformed through successive arithmetic operations that make up the solver algorithm. An interim position where dither is added at various points  within the algorithm may be of interest, but if added on too many operations it will become as expensive as the full SR solution or potentially more so because each Gaussian variate currently used for dithering is more difficult to generate than a uniform variate.

Recent work in large scale climate simulation has explored something analogous to dither in our terms by introducing a noise source into the computational equations \cite{Palmer2015}.

\section{Discussion, further work and conclusion}

In this paper we addressed the numerical accuracy of ODE solvers, solving a well known neuron model in fixed- and floating-point arithmetics.
First, we identified that the constants in the Izhikevich neuron model should be specified explicitly by using the nearest representable number as the GCC fixed-point implementation does round-down in decimal to fixed-point conversion by default (this was also independently noticed by another study \cite{10.3389/fninf.2018.00081} but authors there chose to increase precision of the numerical format of the constants and instead of rounding the constants to the nearest representable value as we did in this work).
Next, we put all constants smaller than 1 into \textit{unsigned long fract} types and developed mixed-format multiplications, instead of keeping everything in \textit{accum}, to maximize the accuracy.
This has not been done in any of the previous studies.
We then identified that fixed-point multiplications are the main remaining source of arithmetic error and explored different rounding schemes.
Round-to-nearest and stochastic rounding on multiplication results are shown to produce substantial accuracy improvements across four ODE solvers and two neuron types.
Fixed-point with stochastic rounding was shown to perform better than fixed-point RN and float, and the mean behaviour is very close to double-precision ODE solvers in terms of spike times. We also found that simple PRNGs will often be good enough for SR to perform well.
The minimum number of bits required in the random number in SR was found to be 6 across four different ODE solvers and two neuron types tested.
In these cases, using more bits is unnecessary, and using fewer will cause the neuron timing to lead compared to the reference.
16-bit arithmetic results were shown to have advantages with SR: although the absolute performance is quite poor (most likely due to overflows and underflows), 16-bit results with SR perform better than 16-bit RD (which largely results in spike time lead) and RN (which produces no spikes).
Further work using scaled interim variables to ameliorate these issues is likely to provide further gains.

We investigated a method of adding noise to the inputs of the ODE on each integration step.
Various levels of noise on the DC current input (dither) were shown to improve the accuracy of many of the solvers and is an alternative way, where high computational performance is a priority, to achieve some of the accuracy improvements that are shown with SR.
Furthermore, we found that stochastic rounding of all the constants, including the fixed timestep value, in each ODE integration step also improves spike timings in most of the test cases.

While the speed of ODE solvers is outside the scope of this paper, measurements in the test bench show that when SR is applied on multiplications, the speed is dictated by \hl{that of the random number generator}.
Preliminary numbers from RK2 Midpoint ODE solver performance benchmarks show an overhead of approximately \SI{30}{\percent} when RN is replaced with SR.
Furthermore, SR is $\sim2.6\times$ faster than software floating-point and $\sim4.2\times$ faster than software double-precision in running a single ODE integration step.

The next generation SpiNNaker chip will have the KISS PRNG in hardware with new random numbers available in one clock cycle, so the overhead of SR compared to RN will be negligible.
With regard to SpiNNaker-2, while it will have a single precision floating-point unit, the results in this paper demonstrate that fixed-point with SR can be more accurate and should be considered instead of simply choosing floating point in a given application.

In the future, we aim to perform mathematical analysis to understand better why SR is causing less rounding error in ODEs as well as to understand in more detail how dither is improving accuracy in the majority of cases.
For performance, we will explore how many random numbers are needed in each ODE integration step: do we require all multipliers to use a separate random number, or is it enough to use one per integration step?
For SpiNNaker neuromorphic applications, we plan to build fast arithmetic libraries with SR and measure their overhead compared to default fixed-point arithmetic and various classical ways of rounding.
Also, we plan to investigate the application of SR in neuron models other than Izhikevich's, and ways to solve them, the first example being LIF with current synapses which has a closed-form solution.
Another direction is to investigate fixed-point arithmetic with SR in solving Partial Differential Equations (PDEs) and other iterative algorithms (e.g. in Linear Algebra).
Finally, it would be beneficial to investigate SR in reduced precision floating-point: \textit{float16} and \textit{bfloat16} numerical types which are becoming increasingly common in the machine learning and numerical algorithms communities, for large-scale projects such as weather simulation using PDEs \cite{Palmer2015, Dawson2018}.

Given all of these results, we predict that any reduced precision system solving ODEs and running other similar iterative algorithms will benefit from adding SR or noise on the inputs, but we would like to confirm this across a wider range of algorithms and problems.
We would also like to point out that different arithmetics have different places where SR could be applied.
For example, unlike fixed-point adders, floating-point adders and subtracters need to round when exponents do not match, and SR could be applied there after the addition has taken place (which would require preserving the bottom bits after matching the exponents). Similarly in neural learning, where the computed changes to a weight are often smaller than the lowest representable value of that weight.
To the best of our knowledge, the application of stochastic rounding in solving ODEs has not yet been investigated on any digital arithmetic, and these are the first results demonstrating substantial numerical error reduction in fixed-point arithmetic.

\dataccess{The data and code used to generate the results presented in this paper are available from doi:10.17632/wxx2mf6n5n.1}

\aucontribute{MM found the issue with rounding in GCC, suggested using SR in the Izhikevich neuron ODE solver and implemented fixed-point libraries with SR. MH implemented PRNGs, fixed-point ODE solvers and a test suite for Izhikevich ODE. MM and MH extended the original ODE testsuite, designed experiments and analysed the data. MH identified arithmetic error as the focal point and carried out the probabilistic analysis of the general SR case. MH and SBF introduced the idea of dither. SBF and DL provided supervision. All authors wrote the manuscript and approved for the final submission.}

\competing{SBF is a founder, director and shareholder of Cogniscience Ltd, which owns SpiNNaker IP. MH and DRL are shareholders of Cogniscience Ltd.}

\funding{The design and construction of the SpiNNaker machine was supported by EPSRC (the UK Engineering and Physical Sciences Research Council) under grants EP/D07908X/1 and	EP/G015740/1, in collaboration with the universities of Southampton, Cambridge and Sheffield and with industry partners ARM Ltd, Silistix Ltd and Thales. Ongoing development of the software, including the work reported here, is supported by the EU ICT Flagship Human Brain Project (H2020 785907), in collaboration with many university and industry partners across the EU and beyond. MM is funded by a Kilburn studentship at the School of Computer Science.
}

\ack{Thanks to Nick Higham for a valuable discussion about backward error analysis, Jim Garside for discussions about the number of bits required in the random numbers in stochastic rounding and to the reviewers for useful comments and references (particularly the work of Prof Chaitin-Chatel in\cite{cc96}) which have improved the paper.}

\bibliographystyle{vancouver}
\bibliography{bibliography,biblio}

\begin{thebibliography}{10}

\bibitem{Jouppi:2017:IPA:3079856.3080246}
Jouppi NP, Young C, Patil N, Patterson D, Agrawal G, Bajwa R, et~al.
\newblock In-Datacenter Performance Analysis of a Tensor Processing Unit.
\newblock In: Proceedings of the 44th Annual International Symposium on
  Computer Architecture. ISCA '17. New York, NY, USA: ACM; 2017. p. 1--12.
\newblock Available from: \url{http://doi.acm.org/10.1145/3079856.3080246}.

\bibitem{KabiB2017AnOverflowFree}
Kabi B, Sahadevan AS, Pradhan T.
\newblock An Overflow Free Fixed-point Eigenvalue Decomposition Algorithm: Case
  Study of Dimensionality Reduction in Hyperspectral Images.
\newblock In: Conference On Design And Architectures For Signal And Image
  Processing (DASIP); 2017. Available from:
  \url{http://dasip2017.esit.rub.de/program.html}.

\bibitem{Gustafson:2017:BFP:3148214.3148220}
Gustafson J, Yonemoto I.
\newblock {Beating Floating Point at Its Own Game: Posit Arithmetic}.
\newblock Supercomput Front Innov: Int J. 2017 Jun;4(2):71--86.

\bibitem{1671767}
{Morris} R.
\newblock Tapered Floating Point: A New Floating-Point Representation.
\newblock IEEE Transactions on Computers. 1971 Dec;C-20(12):1578--1579.

\bibitem{intel_bfloat16}
Intel. {BFLOAT16 -- Hardware Numerics Definition}; 2018.
\newblock Online:
  \url{https://software.intel.com/sites/default/files/managed/40/8b/bf16-hardware-numerics-definition-white-paper.pdf}.

\bibitem{koster2017flexpoint}
K{\"o}ster U, Webb T, Wang X, Nassar M, Bansal AK, Constable W, et~al.
\newblock Flexpoint: An adaptive numerical format for efficient training of
  deep neural networks.
\newblock In: Advances in neural information processing systems; 2017. p.
  1742--1752.
\newblock Available from: \url{https://arxiv.org/abs/1711.02213}.

\bibitem{johnson2018rethinking}
Johnson J.
\newblock Rethinking floating point for deep learning.
\newblock arXiv preprint arXiv:181101721. 2018;Available from:
  \url{https://arxiv.org/abs/1811.01721}.

\bibitem{6757323}
Horowitz M.
\newblock 1.1 Computing's energy problem (and what we can do about it).
\newblock In: 2014 IEEE International Solid-State Circuits Conference Digest of
  Technical Papers (ISSCC); 2014. p. 10--14.

\bibitem{hw-machine-learning-reduced-precision-slides}
Dally W. {High-Performance Hardware for Machine Learning}; 2016.
\newblock Online:
  \url{https://berkeley-deep-learning.github.io/cs294-dl-f16/slides/DL_HW_Berkeley_0916.pdf}.

\bibitem{micikevicius-mixed-precision}
Micikevicius P, Alben J, Garcia D, Ginsburg B, Houston M, Kuchaiev O, et~al.
\newblock {Mixed Precision Training}.
\newblock In: Proceedings of the 6th International Conference on Learning
  Representations. ICLR'18; 2018. .

\bibitem{Gupta:2015:DLL:3045118.3045303}
Gupta S, Agrawal A, Gopalakrishnan K, Narayanan P.
\newblock Deep Learning with Limited Numerical Precision.
\newblock In: Proceedings of the 32nd International Conference on International
  Conference on Machine Learning - Volume 37. ICML'15. JMLR.org; 2015. p.
  1737--1746.
\newblock Available from:
  \url{http://dl.acm.org/citation.cfm?id=3045118.3045303}.

\bibitem{6569370}
Han J, Orshansky M.
\newblock {Approximate computing: An emerging paradigm for energy-efficient
  design}.
\newblock In: 2013 18th IEEE European Test Symposium (ETS); 2013. p. 1--6.

\bibitem{vanderkooy1987dither}
Vanderkooy J, Lipshitz SP.
\newblock Dither in Digital Audio.
\newblock J Audio Eng Soc. 1987;35(12):966--975.
\newblock Available from: \url{http://www.aes.org/e-lib/browse.cfm?elib=5173}.

\bibitem{vanderkooy1989digital}
Vanderkooy J, Lipshitz SP.
\newblock Digital Dither: Processing with Resolution Far Below the Least
  Significant Bit.
\newblock In: Audio Engineering Society Conference: 7th International
  Conference: Audio in Digital Times; 1989. Available from:
  \url{http://www.aes.org/e-lib/browse.cfm?elib=5482}.

\bibitem{6386754}
{Miao} J, {He} K, {Gerstlauer} A, {Orshansky} M.
\newblock Modeling and synthesis of quality-energy optimal approximate adders.
\newblock In: 2012 IEEE/ACM International Conference on Computer-Aided Design
  (ICCAD); 2012. p. 728--735.

\bibitem{spinnproject}
Furber SB, Galluppi F, Temple S, Plana LA.
\newblock {The SpiNNaker Project}.
\newblock Proceedings of the IEEE. 2014 May;102(5):652--665.

\bibitem{10.3389/fnins.2018.00816}
Rhodes O, Bogdan PA, Brenninkmeijer C, Davidson S, Fellows D, Gait A, et~al.
\newblock {sPyNNaker: A Software Package for Running PyNN Simulations on
  SpiNNaker}.
\newblock Frontiers in Neuroscience. 2018;12:816.
\newblock Available from:
  \url{https://www.frontiersin.org/article/10.3389/fnins.2018.00816}.

\bibitem{Hopkins2015}
Hopkins M, Furber S.
\newblock Accuracy and Efficiency in Fixed-Point Neural {ODE} Solvers.
\newblock Neural Computation. 2015;27:2148--2182.

\bibitem{izhikevic2003simple}
Izhikevich EM.
\newblock Simple Model of Spiking Neurons.
\newblock IEEE Transactions on Neural Networks. 2003;14.

\bibitem{iso18037}
ISO/IEC.
\newblock {ISO/IEC, Programming languages - C - Extensions to support embedded
  processors, ISO/IEC TR 18037:2008}.
\newblock ISO/IEC JTC 1/SC 22. 2008;Available from:
  \url{https://www.iso.org/standard/51126.html}.

\bibitem{HOHFELD1992291}
H{\"{o}}hfeld M, Fahlman SE.
\newblock Probabilistic rounding in neural network learning with limited
  precision.
\newblock Neurocomputing. 1992;4(6):291 -- 299.
\newblock Available from:
  \url{http://www.sciencedirect.com/science/article/pii/092523129290014G}.

\bibitem{MullerEtAl2018}
Muller JM, Brunie N, de~Dinechin F, Jeannerod CP, Joldes M, Lef{\`e}vre V,
  et~al.
\newblock Handbook of Floating-Point Arithmetic, 2nd edition.
\newblock {B}irkh\"auser {B}oston; 2018.
\newblock {ACM} {G}.1.0; {G}.1.2; {G}.4; {B}.2.0; {B}.2.4; {F}.2.1., ISBN
  978-3-319-76525-9.

\bibitem{10.3389/fninf.2018.00081}
Trensch G, Gutzen R, Blundell I, Denker M, Morrison A.
\newblock {Rigorous Neural Network Simulations: A Model Substantiation
  Methodology for Increasing the Correctness of Simulation Results in the
  Absence of Experimental Validation Data}.
\newblock Frontiers in Neuroinformatics. 2018;12(81).
\newblock Available from:
  \url{https://www.frontiersin.org/article/10.3389/fninf.2018.00081}.

\bibitem{doi:10.1137/1001011}
Forsythe G.
\newblock Reprint of a Note on Rounding-Off Errors.
\newblock SIAM Review. 1959;1(1):66--67.
\newblock Available from: \url{https://doi.org/10.1137/1001011}.

\bibitem{sjdc07}
Scott NS, J{\'e}z{\'e}quel F, Denis C, Chesneaux JM.
\newblock Numerical `Health Check' for Scientific Codes: {The CADNA} Approach.
\newblock Computer Physics Communications. 2007;176(8):507--521.

\bibitem{alvi08}
Alt R, Vignes J.
\newblock In: Pedrycz W, Skowron A, Kreinovich V, editors. Stochastic
  Arithmetic as a Model of Granular Computing. New York: Wiley; 2008. p.
  33--54.

\bibitem{Higham:2002:ASN}
Higham NJ.
\newblock Accuracy and Stability of Numerical Algorithms.
\newblock 2nd ed. Philadelphia, PA, USA: Society for Industrial and Applied
  Mathematics; 2002.

\bibitem{Muller2015}
M\"{u}ller LK, Indiveri G.
\newblock {Rounding Methods for Neural Networks with Low Resolution Synaptic
  Weights}.
\newblock arXiv. 2015;Available from: \url{http://arxiv.org/abs/1504.05767}.

\bibitem{Wang2018}
Wang N, Choi J, Brand D, Chen CY, Gopalakrishnan K.
\newblock {Training Deep Neural Networks with 8-bit Floating Point Numbers}.
\newblock In: NeurIPS; 2018. .

\bibitem{10.3389/fnins.2018.00745}
Gokmen T, Rasch MJ, Haensch W.
\newblock {Training LSTM Networks With Resistive Cross-Point Devices}.
\newblock Frontiers in Neuroscience. 2018;12:745.
\newblock Available from:
  \url{https://www.frontiersin.org/article/10.3389/fnins.2018.00745}.

\bibitem{8259423}
Davies M, Srinivasa N, Lin TH, Chinya G, Cao Y, Choday SH, et~al.
\newblock {Loihi: A Neuromorphic Manycore Processor with On-Chip Learning}.
\newblock IEEE Micro. 2018 January;38(1):82--99.

\bibitem{marsKISS64}
Marsaglia G, Zaman A.
\newblock The KISS generator.
\newblock Department of Statistics, Florida State University, Tallahassee, FL,
  USA.; 1993.

\bibitem{2007TestU01}
L'Ecuyer P, Simard R.
\newblock {TestU01: A C library for empirical testing of random number
  generators}.
\newblock ACM Trans Math Softw. 2007;33(4).

\bibitem{dieharder-weblink}
Brown R. dieharder;.
\newblock Available from:
  \url{http://webhome.phy.duke.edu/~rgb/General/dieharder.php}.

\bibitem{Golomb:1981:SRS:578271}
Golomb SW.
\newblock Shift Register Sequences.
\newblock Laguna Hills, CA, USA: Aegean Park Press; 1981.

\bibitem{press1992numerical}
Press WH, Teukolsky SA, Vetterling WT, Flannery BP.
\newblock Numerical {Recipes} in {C} (2nd ed.).
\newblock Cambridge; 1992.

\bibitem{hall1976modern}
Hall G, Watt JM, editors.
\newblock Modern Numerical Methods for Ordinary Differential Equations.
\newblock Clarendon press; 1976.

\bibitem{lambert1991numerical}
Lambert JD.
\newblock Numerical Methods for Ordinary Differential Systems: The Initial
  Value Problem.
\newblock Wiley; 1991.

\bibitem{dormand1996numerical}
Dormand JR.
\newblock Numerical Methods for Differential Equations.
\newblock CRC press; 1996.

\bibitem{butcher2003numerical}
Butcher JC.
\newblock Numerical Methods for Ordinary Differential Equations.
\newblock Wiley; 2003.

\bibitem{research2014mathematica}
{Wolfram Research}.
\newblock Mathematica.
\newblock Version 10.0 ed. Champaign, Illinois, {USA}: Wolfram Research, Inc.;
  2014.

\bibitem{Chan:2010aa}
Chan RPK, Tsai AYJ.
\newblock {On explicit two-derivative Runge-Kutta methods}
  [10.1007/s11075-009-9349-1].
\newblock Numerical Algorithms. 2010;53(2-3):171--194.
\newblock Available from: \url{http://dx.doi.org/10.1007/s11075-009-9349-1}.

\bibitem{cc96}
Chaitin-Chatelin F, Frayss{\'e} V.
\newblock Lectures on Finite Precision Computations.
\newblock SIAM; 1996.

\bibitem{Palmer2015}
Palmer T.
\newblock {Build imprecise supercomputers}.
\newblock Nature. 2015;526:32--33.

\bibitem{Dawson2018}
Dawson A, D{\"u}ben PD, MacLeod DA, Palmer TN.
\newblock Reliable low precision simulations in land surface models.
\newblock Climate Dynamics. 2018 Oct;51(7):2657--2666.
\newblock Available from: \url{https://doi.org/10.1007/s00382-017-4034-x}.

\end{thebibliography}

\end{document}